\documentclass[cernpreprint, atlasdraft=true, texlive=2016, UKenglish, texmf]{atlasdoc}  
 
\usepackage{atlaspackage}
\usepackage{hyperref}
 
\usepackage{multirow}
 
\usepackage{atlasbiblatex}
 
\usepackage{atlascontribute}
 
\usepackage{atlasphysics}
 
\graphicspath{{logos/}{figures/}}

\AtlasTitle{Performance of the ATLAS Level-1 topological trigger in Run~2}
 
\PreprintIdNumber{CERN-EP-2021-40}
 
\AtlasAbstract{
During LHC Run~2 (2015--2018) the ATLAS Level-1 topological trigger
allowed efficient data-taking by the ATLAS experiment at luminosities up to 2.1$\times$10$^{34}$~cm$^{-2}$s$^{-1}$,
which exceeds the design value by a factor of two.
The system was installed in 2016 and operated in 2017 and 2018. It uses Field Programmable Gate Array processors to select interesting events by placing kinematic and angular requirements on electromagnetic clusters, jets, $\tau$-leptons, muons and the missing transverse energy.
It allowed to significantly improve the background event rejection and signal event acceptance, in particular for Higgs and $B$-physics processes.}
 
\author{The ATLAS Collaboration}

\AtlasRefCode{TRIG-2019-02}
 
\AtlasJournalRef{Eur. Phys. J. C 82 (2022) 1}   
\AtlasDOI{10.1140/epjc/s10052-021-09807-0}
 
\AtlasNote{TRIG-2019-02}

\hypersetup{pdftitle={ATLAS document},pdfauthor={The ATLAS Collaboration}}
 
\begin{document}
 
\maketitle
 
\tableofcontents
 
\clearpage
 
\FloatBarrier
\section{Introduction}
During Run~2, the Large Hadron Collider (LHC) delivered proton--proton ($pp$) collisions to the ATLAS experiment~\cite{PERF-2007-01} with instantaneous luminosities up to 2.1$\times$10$^{34}$~cm$^{-2}$s$^{-1}$ and a centre-of-mass energy of 13~\TeV. The very high instantaneous luminosity, accompanied by unprecedented rates of simultaneous $pp$ interactions (pile-up) per bunch crossing (BC), placed stringent operational constraints~\cite{Aad:2020wji} on the ATLAS trigger system~\cite{Aaboud:2016leb} to efficiently select the collision events relevant for the ATLAS physics programme. Events from the 40~MHz BCs were accepted at a reduced rate compatible with the detector readout, storage and offline computing resources, about 1--1.5~kHz on average.
 
\begin{figure}[tbh]
\begin{center}
\includegraphics[width=0.8\textwidth]{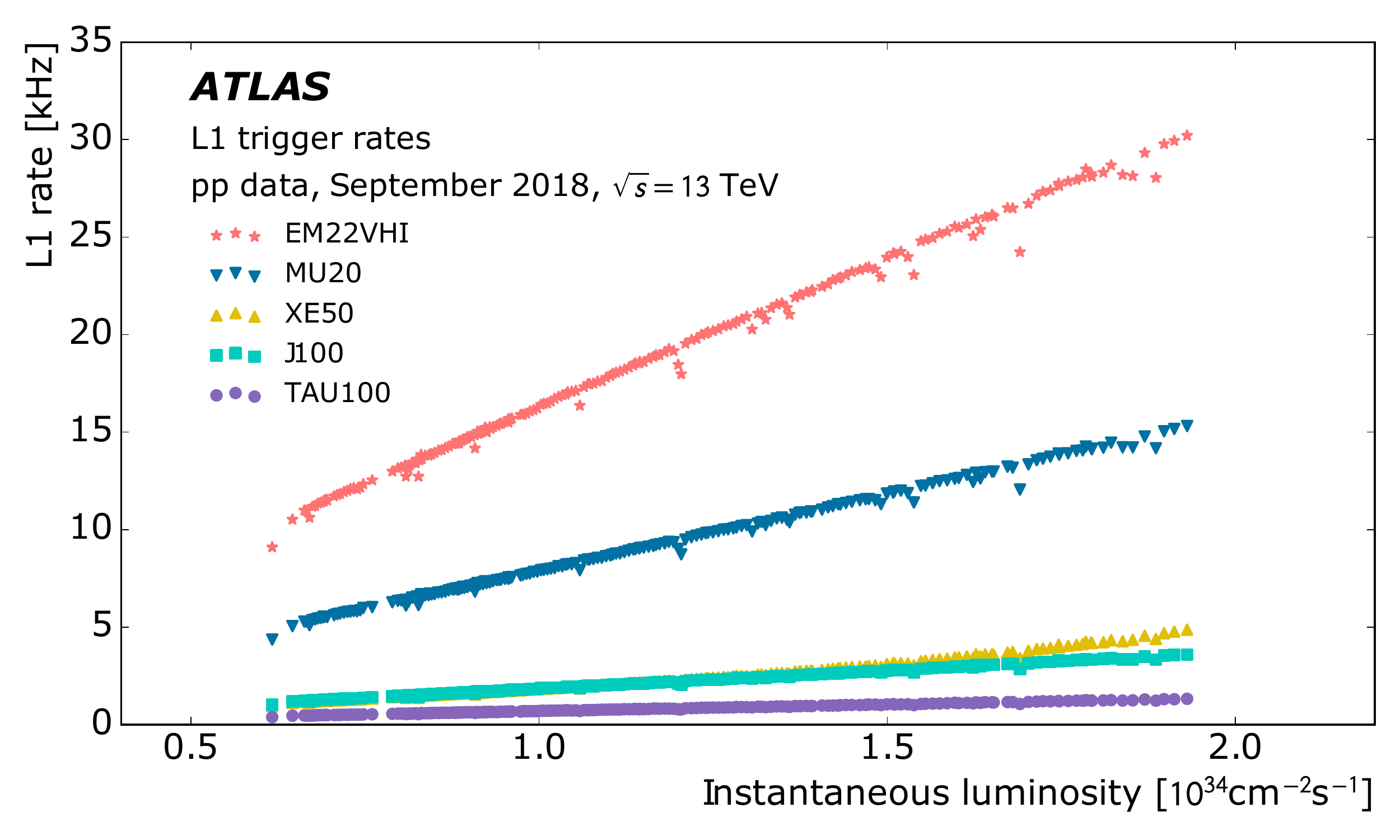}
\end{center}
\caption{
Event rate of selected L1 trigger items as a function of the instantaneous luminosity in an LHC fill taken in September 2018 with a peak luminosity of 2.0~$\times$~10$^{34}$~cm$^{-2}$s$^{-1}$ and a peak average number of interactions per BC of 56. Presented are rates of some representative single-object trigger items that have not been prescaled. These trigger items are based on such objects as electromagnetic clusters (EM), muon candidates (MU), jet candidates (J), missing transverse energy (XE) and $\tau$-lepton candidates (TAU). The number in the trigger name denotes the trigger transverse momentum threshold in \GeV. The letters following the threshold value in the EM22VHI trigger refer to details of the selection: variable thresholds (V), hadronic isolation (H), and electromagnetic isolation (I). }
\label{fig:rate_vs_lumi}
\end{figure}

The Level-1 (L1) trigger is the first rate-reducing step in the ATLAS trigger system.
As shown in Figure \ref{fig:rate_vs_lumi}, the rate of events selected by some L1 physics triggers, named trigger `items' in this paper, increases with the LHC instantaneous luminosity. To cope with such a rate increase, three approaches were used previously:
tightening the requirements on the selected trigger objects, with a potential loss in signal acceptance;
recording only one event in every $N$ events (trigger prescale), with a consequent loss of integrated luminosity;
or combining different trigger objects.
 
All three approaches cause some loss of interesting data, and this is particularly unfavourable for physics
signatures having low-momentum final-state particles.
For the first time in ATLAS, the L1 topological trigger (L1Topo) provided the ability to implement topological requirements, i.e.\ criteria
based on the kinematic relations between objects, to improve the rejection of background events
while preserving high acceptance for signal events.
 
This paper describes the performance of the L1Topo trigger system during the data-taking in 2017 and 2018.
The paper is organised as follows. The ATLAS detector and the architecture of the trigger system are described in Section~\ref{sec:detector}. An overview of the topological processor and its algorithms is presented in Sections~\ref{sec:hardware} and \ref{sec:algorithms}. The results of the validation procedure for such algorithms and their performance in physics events are discussed in Sections~\ref{sec:validation} and \ref{sec:performance}, followed by the conclusions in Section~\ref{sec:conclusion}.
 
 
\FloatBarrier
\section{The ATLAS detector and trigger architecture}
\label{sec:detector}
\newcommand{\AtlasCoordFootnote}{
ATLAS uses a right-handed coordinate system with its origin at the nominal interaction point (IP)
in the centre of the detector and the $z$-axis along the beam pipe.
The $x$-axis points from the IP to the centre of the LHC ring,
and the $y$-axis points upwards.
Cylindrical coordinates $(r,\phi)$ are used in the transverse plane,
$\phi$ being the azimuthal angle around the $z$-axis.
The pseudorapidity is defined in terms of the polar angle $\theta$ as $\eta = -\ln \tan(\theta/2)$.}

The ATLAS experiment~\cite{ATLAS:2008xda} at the LHC is a multipurpose particle detector
with a forward--backward symmetric cylindrical geometry and a near $4\pi$ coverage in
solid angle.\footnote{\AtlasCoordFootnote}
It consists of an inner tracking detector surrounded by a thin superconducting solenoid
providing a \SI{2}{\tesla} axial magnetic field, electromagnetic and hadronic calorimeters, and a muon spectrometer.
The inner tracking detector covers the pseudorapidity range $|\eta| < 2.5$.
It consists of silicon pixel, silicon microstrip, and transition-radiation tracking detectors.
Lead/liquid-argon sampling calorimeters provide electromagnetic energy measurements
with high granularity for $|\eta| < 2.5$.
A steel/scintillator-tile hadron calorimeter covers the central pseudorapidity range ($|\eta| < 1.7$).
The endcap and forward regions are instrumented with calorimeters based on liquid argon technology
for both the electromagnetic (EM) and hadronic energy measurements up to $|\eta| = 4.9$.
The muon spectrometer surrounds the calorimeters and is based on
three large air-core toroidal superconducting magnets with eight coils each.
The field integral of the toroids ranges between \num{2.0} and 6.0 Tm
across most of the detector.
The muon spectrometer includes three layers of high-precision tracking chambers which provide coverage in the range $|\eta| < 2.7$,
and fast detectors for triggering in the range $|\eta| < 2.4$.
 
The ATLAS experiment uses a two-level trigger system to select interesting events to record and analyse offline.
The L1 trigger is hardware-based and uses a subset of the detector information to accept events from the 40~MHz LHC $pp$ BCs at a rate below 100~kHz, which is the maximum detector readout rate. This is followed by a software-based high-level trigger (HLT) which reduces the event rate to around 1~kHz on average, constrained by a data storage capability of a few hundred MB/s~\cite{ATLASTDAQ:2016pov}.
 
\begin{figure}[!t]
\centering
\includegraphics[width=0.8\textwidth]{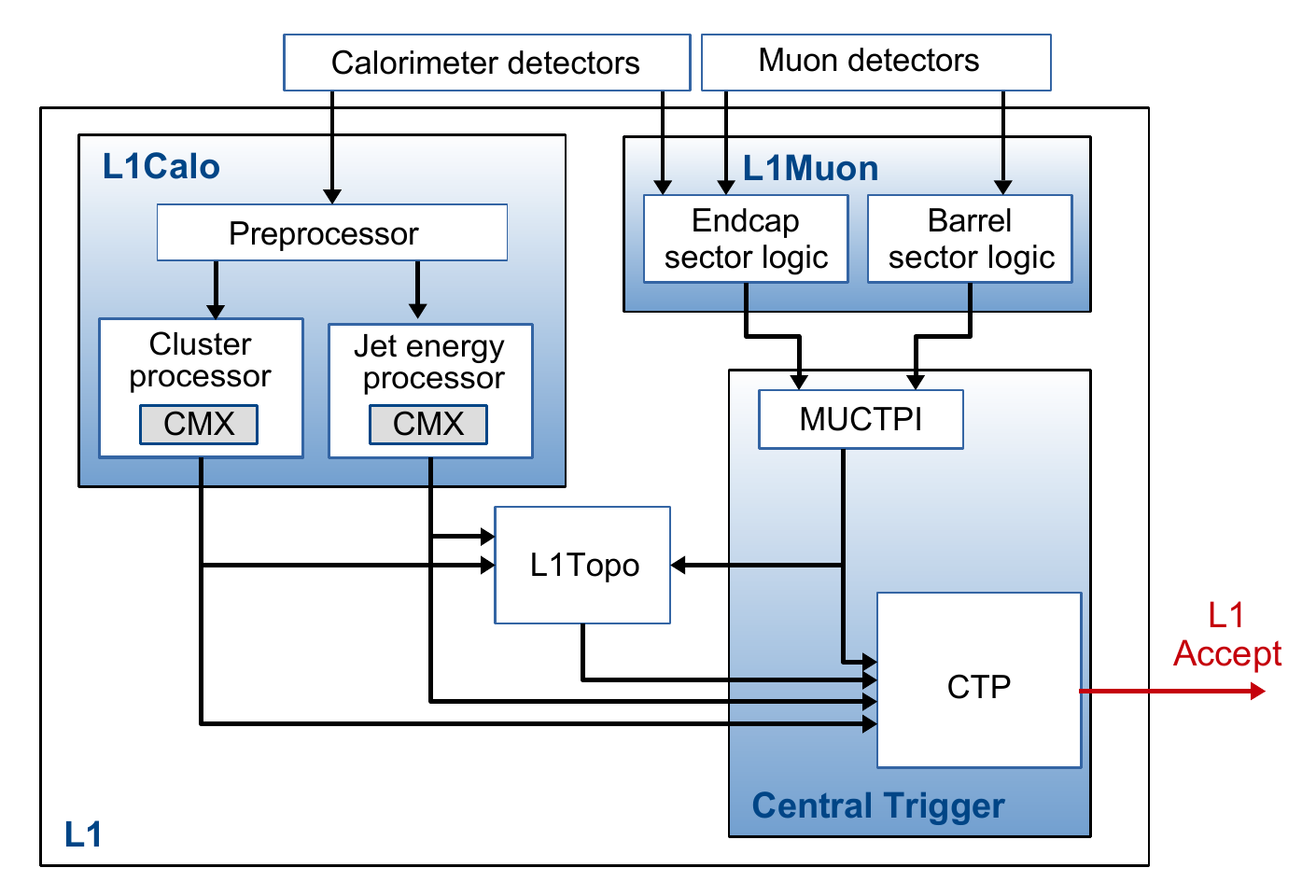}
\caption{The ATLAS L1 trigger system in Run~2~\cite{Aad:2020wji}.}
\label{fig:atlas-trigger}
\end{figure}

The L1 trigger system, sketched in Figure~\ref{fig:atlas-trigger}, consists of the calorimeter trigger (L1Calo)~\cite{Achenbach:2008zzb} and muon trigger (L1Muon)~\cite{Aad:2020uyd}
systems, which compute the multiplicity of calorimeter energy clusters and muons above a
transverse momentum (\pt) threshold (representing the minimum energy required),\footnote{L1Calo triggers compute the multiplicity of clusters above transverse energy (\et) thresholds, while L1Muon triggers compute the multiplicity of muons above transverse momentum (\pt) thresholds.} 
as well as the Central Trigger Processor (CTP), which receives this information and computes the final L1 decision. In 2015 and 2016, new modules were installed in the real-time data path. The L1Calo and L1Muon systems were respectively upgraded with the Common Merger Modules (CMX) and the Muon Interface to L1Topo (MUCTPI2Topo) modules to store the momentum and angle ($\eta$ and $\phi$ coordinate) information
for each candidate object and propagate it to the new L1Topo processor. For the first time in ATLAS, this information was available in a single L1 trigger system and used to compute kinematic quantities and relations among multiple trigger objects, such as angular distances or invariant masses.
 
\FloatBarrier
\section{The L1 topological processor}
\label{sec:hardware}
\begin{figure}[hbt]
\centering
\subfloat{
\includegraphics[height=0.3\textwidth]{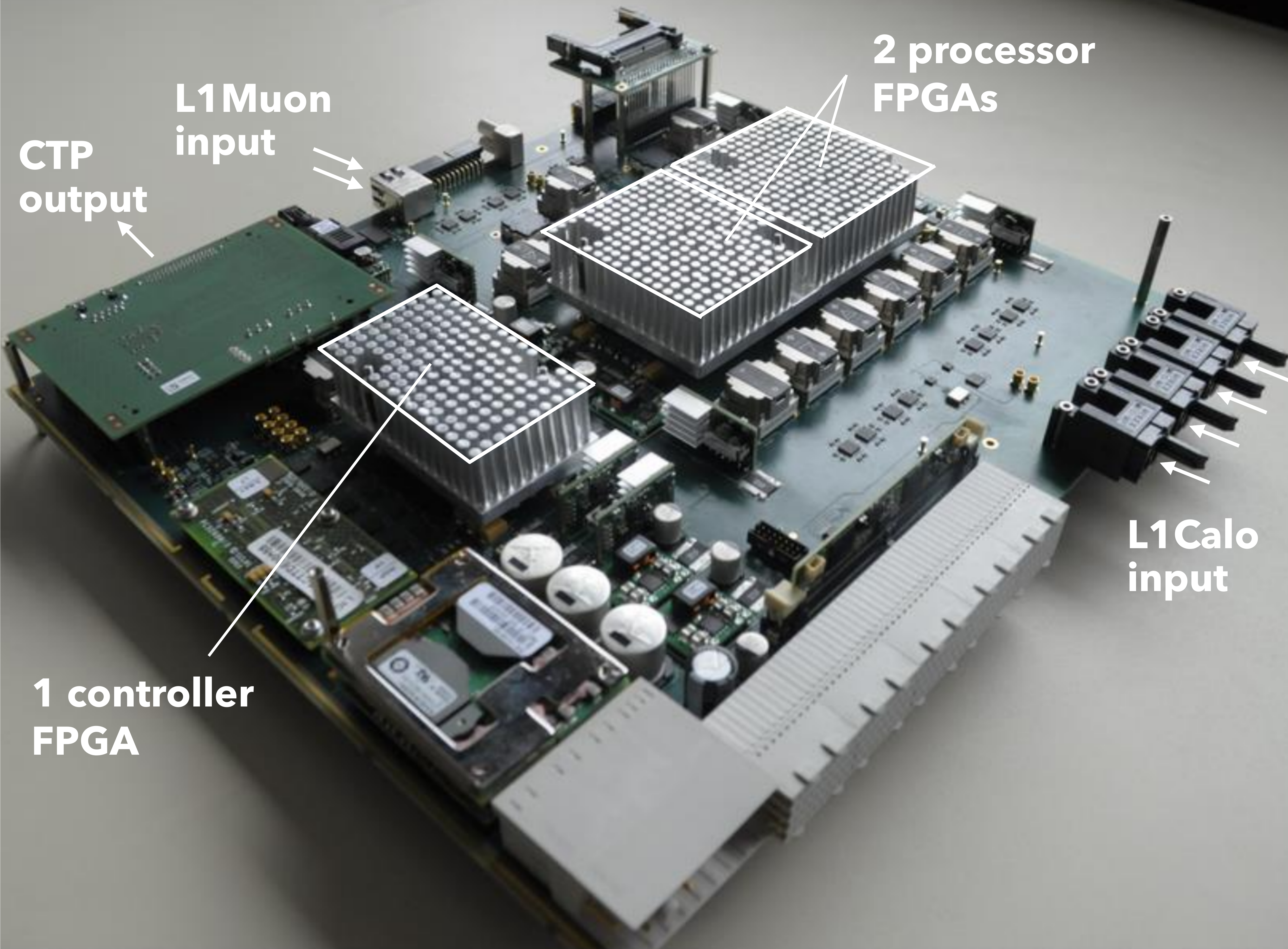}
\label{fig:module}
}
\subfloat{
\includegraphics[height=0.3\textwidth]{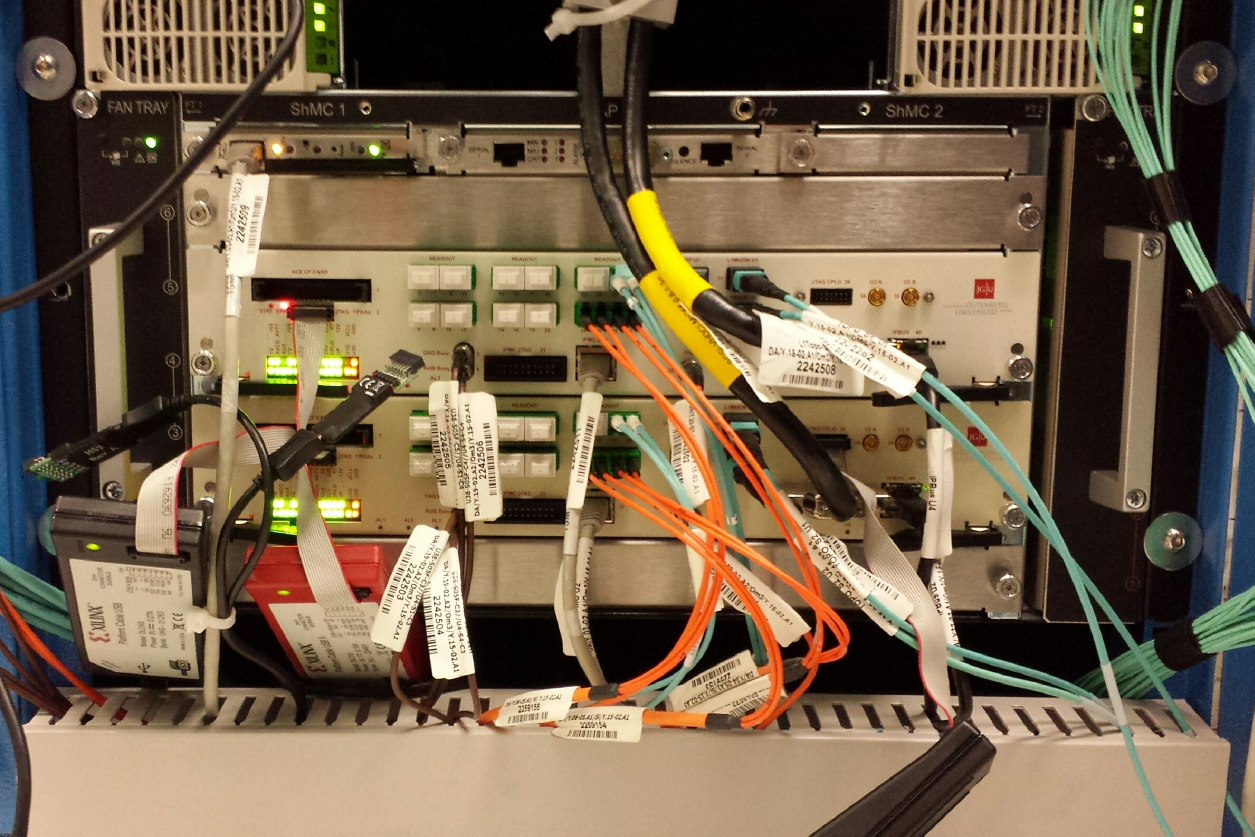}
\label{fig:operation}
}
\caption{\protect\subref{fig:module} One of the L1Topo modules before fibre assembly, \protect\subref{fig:operation} the L1Topo system in operation.}
\label{fig:topo}
\end{figure}

The L1Topo system consists of a single processor crate equipped with two identical AdvancedTCA-compliant modules~\cite{atca}. Pictures of one of the L1Topo modules before fibre assembly and of the full L1Topo system during operation are shown in Figure~\ref{fig:topo}. Each module has two processor Field Programmable Gate Arrays (FPGAs) (Xilinx Virtex7~\cite{Virtex}) to run the algorithms and one controller FPGA (Kintex7~\cite{Kintex}) responsible for the readout and all communication logic to interface to the external trigger systems.
 
\begin{figure}[hbt]
\centering
\includegraphics[width=0.8\textwidth]{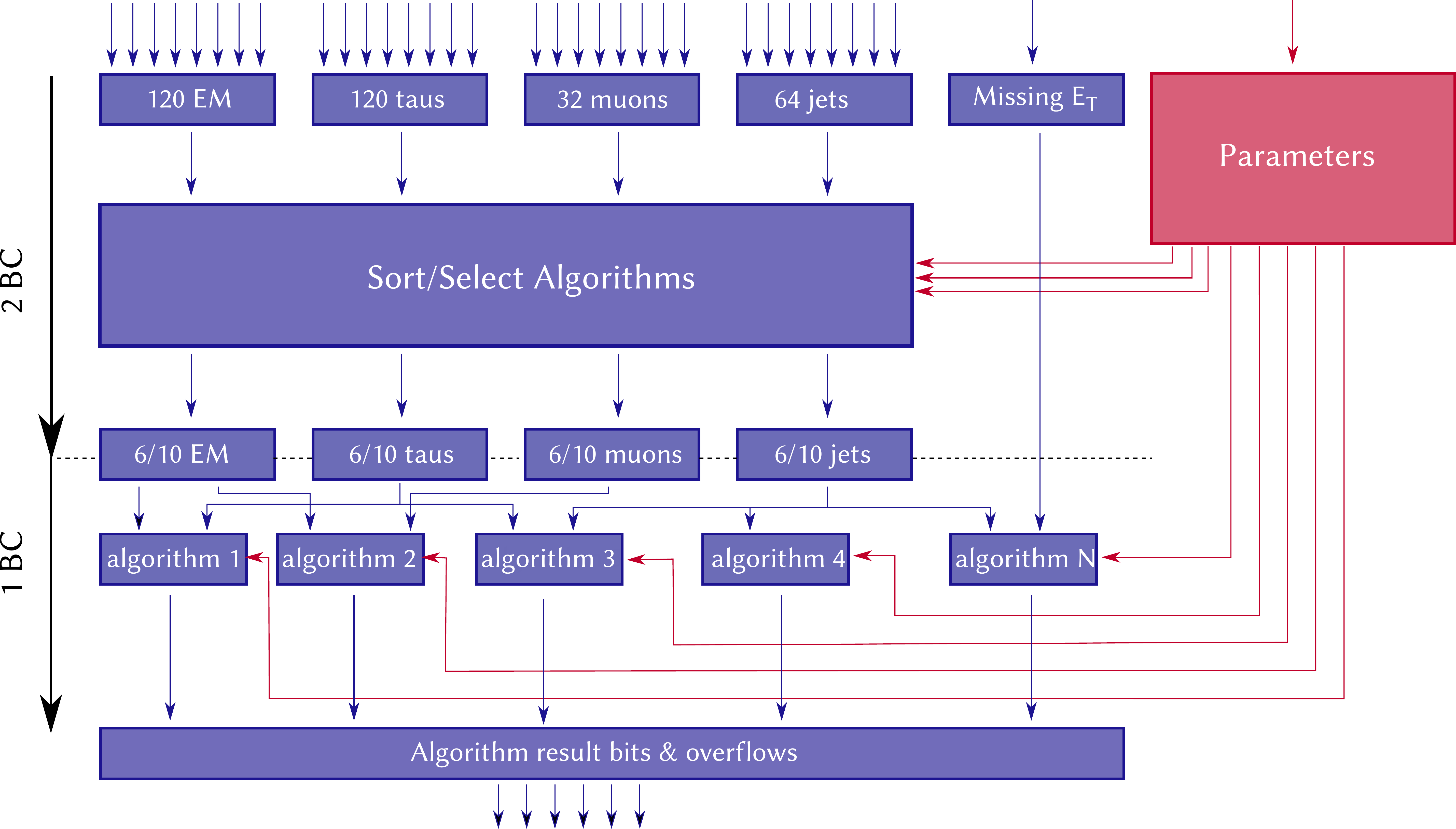}
\caption{Structure of the topological algorithms. The diagram provides a schematic view of the L1Topo data-flow, starting from the full TOBs input list to the output of the topological algorithms. Different kinds of input TOBs can be combined in each algorithm as needed for the physics signature. One LHC BC interval corresponds to 25~ns.}
\label{fig:algorithms}
\end{figure}

A sketch of the architecture of the topological algorithms is shown in Figure~\ref{fig:algorithms}.
The input data are sent via optical fibres at a speed of 6.4~Gb/s per line to the L1Topo system back-plane and front-panel from the L1Calo and L1Muon systems, respectively.
The data format is defined by Trigger Objects (TOBs), which are bit arrays encoding the position ($\eta$, $\phi$), \pT{}, and further qualifying
information\footnote{Examples are the isolation of electromagnetic clusters and $\tau$-leptons, or the jet momentum obtained by clustering the object with a larger window size.}
for each object. The maximum numbers of different TOBs that can be received by the L1Topo system in one event are:
64 jet TOBs, 120 EM TOBs~\cite{Aad:2019wsl}, 120 tau TOBs, 32 muon TOBs~\cite{Aad:2020uyd} and 1 missing transverse energy TOB~\cite{Aad:2020les};
the maximum number of TOBs received in the L1Topo system was found to be well within the constraints imposed by the physics cases of interest for the L1Topo trigger.
As shown in Table~\ref{tab:tobs-gran}, each type of TOB is characterised by a different granularity and bitwise precision for \pt{} and the $|\eta|$ and $\phi$ coordinates.
 
The optical signals are converted into electrical signals via Avago miniPODs~\cite{minipods} and then directed into the processor FPGAs, where they are deserialised in multi-gigabit transceivers.
In each module, both processors are supplied with the same data, so that they can operate independently and in parallel.
For each algorithm the resulting output action consists of setting two bits: the algorithm decision and the overflow bit.
The overflow bit is set if any module of the L1Calo or L1Muon systems identifies more objects than it can provide to L1Topo: 5 (4) clusters (jets) per input L1Calo CMX module, or 2 muons per L1Muon Interface Octant board (MIOCT)~\cite{CERN-LHCC-98-014}, respectively.
The logical OR of the two bits is then transmitted to the CTP for use in the L1 trigger decision.

\begin{table}[htb]
\begin{center}
\caption{Coordinate granularity and range for all the TOBs input to L1Topo.
The $\eta\times\phi$ granularity for EM, tau and jet TOBs depends upon the detector system and readout; it ranges between $0.1\times 0.1$ for EM and tau TOBs to $1.7\times 0.2$ for jet TOBs with $|\eta|\geq3.2$.
The precision of the bitwise representation of each TOB is also shown.
The \pt{} measurement for muon TOBs used in L1Topo corresponds to the three available L1 \pt{} thresholds, and the $\eta$ granularity is irregular ranging between $0.2$ and $0.4$.}
\vspace{0.2cm}
\begin{tabular}{lccccccc}
\hline
\cline{1-8}
\multicolumn{1}{l|}{\multirow{2}{*}{TOB type}} & \multicolumn{3}{c}{\multirow{2}{*}{Coordinate granularity}} & \multirow{2}{*}{$|\eta|$ range} & \multicolumn{3}{c}{Bitwise precision} \\
\multicolumn{1}{l|}{}& \multicolumn{3}{c}{}                       &                                          & \multicolumn{3}{c}{(number of bits)} \\
\hline
\cline{1-8}
&  &  &  &  & & &  \\
\cline{2-8}
\multicolumn{1}{l|}{}
& \pt\ [GeV] & $\eta$ & $\phi$ [rad]         &                & \pt\  & $\eta$ & $\phi$   \\
\hline
\multicolumn{1}{l|}{EM and tau}         &  ~~0.5                         &  $0.1$                                   &  $0.1$        &  [0, 2.5]      & 8 & 6 & 6 \\
\multicolumn{1}{l|}{\multirow{2}{*}{Jets}}      &  1~                            &  $0.2$                                   &  $0.2$        &  [0, 3.2]      & \multirow{2}{*}{10} & \multirow{2}{*}{5} & \multirow{2}{*}{5} \\
\multicolumn{1}{l|}{}   &  1~                            &  $1.7$                                   &  $0.2$        &  [3.2, 4.9]~~~ & &  \\
\hline
&  &  &  &  & & & \\
\cline{2-8}
\multicolumn{1}{l|}{} & \pt\ [GeV] & \multirow{2}{*}{$\eta$} & \multirow{2}{*}{$\phi$ [rad]}         &                & \multirow{2}{*}{\pt}  & \multirow{2}{*}{$\eta$} & \multirow{2}{*}{$\phi$} \\
\multicolumn{1}{l|}{} &  (thresholds) &  &  &  & & &  \\
\hline
\multicolumn{1}{l|}{Muons}              &  $\{4, 6, 10 \}$   &  $0.2$--$0.4$                            &  $0.1$        &  [0, 2.4]      & 2 & 6 & 6 \\
\hline
&  &  &  &  & & &  \\
\cline{2-8}
\multicolumn{1}{l|}{}               & E$_{\mathrm{x}}$ [GeV] & E$_{\mathrm{y}}$ [GeV] & $\phi$ [rad]      &                & E$_{\mathrm{x}}$ & E$_{\mathrm{y}}$ & $\phi$  \\
\hline
\multicolumn{1}{l|}{\met{}}   & $1$                     & $1$                    & $0.1$         & -              & $16$            & $16$           & 6  \\
\hline
 
\end{tabular}
\label{tab:tobs-gran}
\end{center}
\end{table}
 
\begin{table}[tbh]
\begin{center}
\caption{
Resources used for lookup table blocks (Configurable Logic Blocks~\cite{Virtex}) and number of output bits per processor FPGA out of a maximum of 32.
} \label{tab:resourceusage}
\vspace{0.2cm}
\begin{tabular}{l|c|c|c|c}
\hline
\textbf{Processor FPGA} & \textbf{1} & \textbf{2}& \textbf{3}& \textbf{4}\\
\hline
Percentage of lookup table blocks utilised  & 50\% & 58\% &71\% & 71\% \\
Number of output bits & 32~~ & 23~~ & 27~~ & 31~~ \\
\hline
\end{tabular}
\end{center}
\end{table}
 
Thanks to the extensive logic resources of the processor FPGAs, up to 128 topological algorithms (32 per processor FPGA) can be executed on the real-time data path and are configured using a set of parameters. By the end of 2018, a total of 113 topological algorithms were implemented in VHDL~\cite{vhdl}, validated and operated. Lookup table operations dominate the resource usage in the FPGAs, as shown in Table~\ref{tab:resourceusage} per FPGA together with the number of output bits.
 
\begin{figure}[htb]
\centering
\includegraphics[width=0.8\textwidth]{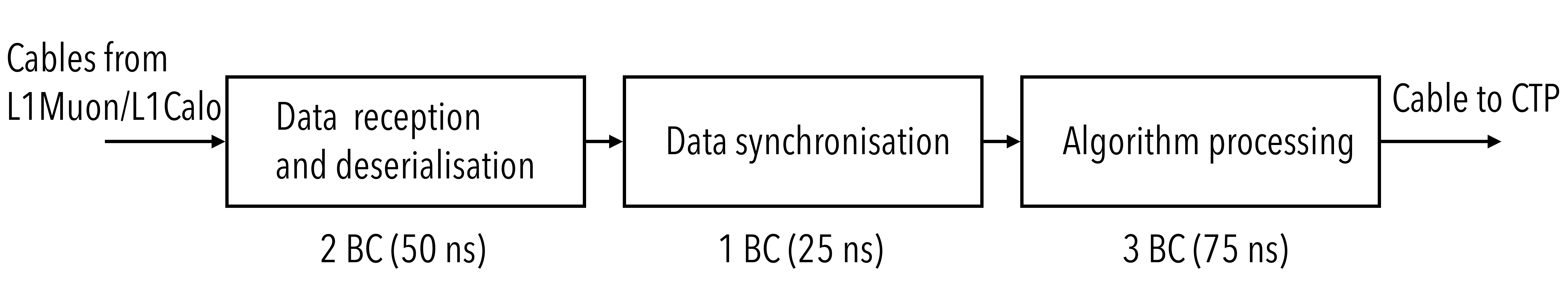}
\caption{Latency of the L1Topo system. The total latency is $\sim$200~ns: 150~ns account for the operation of the system as shown in the figure, and the remaining 50~ns account for the length of the input and output cables.}
\label{fig:latency}
\end{figure}

The total latency of the L1Topo system is $\sim$200~ns, corresponding to eight LHC BCs of 25~ns. As summarised in Figure~\ref{fig:latency},
50~ns are used to receive and deserialise the data, 25~ns for their synchronisation and 75~ns are dedicated to the execution of the topological algorithms.
The remaining 50~ns accounts for the data transmission through cables from the input source channels to L1Topo and from L1Topo to the final destination.
Regarding the topological algorithms, the first 50~ns are used to filter the input TOBs in order
to reduce the number of possible combinations when considering relations among objects, thus reducing the
resource usage per algorithm. Technically, this is achieved by creating two types of filtered lists using the so-called `sort/select' algorithms in Figure~\ref{fig:algorithms}. The first list type is created by sorting the TOBs by \pT{} and taking the first six leading objects. 
The second list type is created by selecting the first ten objects above a \pT{} threshold as ordered in the input source channel and received in the FPGA. 
An overflow bit is set for events with a second list type of more than ten TOBs. 
The length of these filtered lists is mostly limited by latency constraints, and it is adequate for most physics signatures.
Finally, the remaining 25~ns in the algorithm execution are used to run the decision algorithms on
the desired TOB lists. During Run~2 operations, a total of 17 lists including the two types were considered for the whole L1Topo system.
 
The set of criteria, the corresponding threshold values, and the lists of input objects are read from a programmable trigger menu~\cite{Aad:2020wji} for each algorithm.
 
\newpage
 
\FloatBarrier
\section{The topological algorithms}
\label{sec:algorithms}
This section provides a general overview of the topological algorithms and their scope, while more details of specific use-cases are discussed in Section \ref{sec:performance}.
 
Different categories of topological algorithms have been implemented in L1Topo to address the needs of physics analyses and to help in the commissioning of new systems of the ATLAS detector.  A complete list of topological algorithms is detailed in Table~\ref{tab:algo_def}. The exact L1 trigger object input(s) to the algorithms, their specific requirements in terms of \pt, and the exact selection cuts applied by the algorithms are configurable and defined in the ATLAS trigger menu~\cite{ATL-DAQ-PUB-2017-001,ATL-DAQ-PUB-2018-002,ATL-DAQ-PUB-2019-001}. The trigger menu configuration can be changed during runtime, while the topological algorithms are defined in L1Topo firmware and cannot be modified during the data-taking without remotely accessing the L1Topo boards.
 
\begin{table}[!htbp]
\caption{List of topological algorithms implemented in Run~2. The input L1 trigger objects are required to pass configurable \pt{} threshold, $\eta$ and isolation requirements. Trigonometric functions in the FPGAs are calculated using lookup tables.}
\vspace{0.2cm}
\begin{center}
\setlength{\tabcolsep}{0.0pc}{\small
 
\begin{tabular*}{\textwidth}{p{7 cm} p{9cm}}
\noalign{\smallskip}\hline\noalign{\smallskip}
Algorithm       & Definition          \\[-0.05cm]
\noalign{\smallskip}\hline\noalign{\smallskip}
 
$\mathrm{Pseudorapidity\ distance}$: $\ \ \ \ \ $ \hspace{150pt} $x_1 \leq \Delta\eta = |\eta_1 - \eta_2| \leq x_2$  & Pseudorapidity difference between the objects of two given input lists.  The lower and upper bounds ($x_1, x_2$) are configurable parameters of the algorithm. \\
\noalign{\smallskip}\hline\noalign{\smallskip}
 
$\mathrm{Azimuthal\ distance}$: $\ \ \ \ $ \hspace{150pt} $y_1 \leq \Delta\phi= |\phi_1 - \phi_2| \leq y_2$ & Azimuthal distance between the objects of two given input lists. The lower and upper bounds ($y_1, y_2$) are configurable parameters of the algorithm.  \\
\noalign{\smallskip}\hline\noalign{\smallskip}
 
$\mathrm{Box\ cuts}$: $\ \ \ \ \ $ \hspace{150pt} $x_1 \leq \Delta\eta \leq x_2$ and $y_1 \leq \Delta\phi \leq y_2$  & Combination of $\Delta\eta$ and $\Delta\phi$ requirements with configurable lower and upper bounds ($x_1, y_1$, $x_2, y_2$). The same pair of objects must satisfy both conditions. \\
\noalign{\smallskip}\hline\noalign{\smallskip}
 
$\mathrm{Window\ cuts}$: $\ \ \ \ \ $ \hspace{150pt} $x_1 < \eta < x_2$ and $y_1< \phi < y_2$  & Requirement on the pseudorapidity and azimuthal positions of the objects given in an input list. The lower and upper bounds ($x_1, y_1$, $x_2, y_2$) are configurable parameters of the algorithm.  \\
\noalign{\smallskip}\hline\noalign{\smallskip}
 
$\mathrm{Angular\ distance}$: $\ \ \ \ \ $ \hspace{150pt} $x_1^2 \leq \Delta R^2=(\Delta\phi)^2+(\Delta\eta)^2 \leq x_2^2$ & Angular distance between the objects of two given input lists. The lower and upper bounds ($x_1, x_2$) are configurable parameters of the algorithm. \\
\noalign{\smallskip}\hline\noalign{\smallskip}
 
Disambiguation: $\ \ \ \ $ \hspace{150pt} $\phi_1 \neq \phi_2 \,\,\, \mathrm{OR} \,\,\, \eta_1 \neq \eta_2$, $\Delta R>x$ & Takes two or three input lists and checks the spatial separation between each pair. For objects with the same granularity the $\phi$ and $\eta$ coordinates are used directly, otherwise a requirement on $\Delta R$ is imposed.    \\
\noalign{\smallskip}\hline\noalign{\smallskip}
 
Ratio: $O_{1}\geq x \cdot O_{2}$ & Takes two input lists and for two objects in the same location ($\phi_1 = \phi_2~\mathrm{AND}~\eta_1 = \eta_2$) the ratio of two quantities is required to be above a threshold. The threshold ($x$) and the two quantities ($O_1, O_2$) are configurable parameters of the algorithm.  \\
 
\noalign{\smallskip}\hline\noalign{\smallskip}
 
$\mathrm{Invariant\ mass}$: $\ \ \ \ $ \hspace{150pt}  $x_1^2\leq m^2_{\mathrm{inv}}=2\,E_{\mathrm{T,1}}E_{\mathrm{T,2}}(\cosh\Delta\eta-\cos\Delta\phi) \leq x_2^2$ & Invariant mass of each pair of objects in two input lists. If the two input lists are different, the two objects used in the mass computation must satisfy the disambiguation criteria. The lower and upper bounds ($x_1, x_2$) are configurable parameters of the algorithm. \\
 
\noalign{\smallskip}\hline\noalign{\smallskip}

$\mathrm{Transverse\ mass}$: $\ \ \ \ $ \hspace{140pt}  $x_1^2 \leq m_{\mathrm{T}}^2=2\,E_{\mathrm{T,1}}E_{\mathrm{T}}^{\mathrm{miss}}(1- \cos\Delta\phi) \leq x_2^2$ & Transverse mass of the \met{} and the objects in one input list. The lower and upper bounds ($x_1, x_2$) are configurable parameters of the algorithm.  \\

\noalign{\smallskip}\hline\noalign{\smallskip}

Event hardness: $H_{\mathrm{T}}>x$ & Scalar sum of the transverse momenta of all or a selected number of jets. Additional requirements on the jet \pt{} or $\eta$ are possible. The threshold ($x$) is a configurable parameter of the algorithm. \\
\noalign{\smallskip}\hline\noalign{\smallskip}

Simple cone: $\displaystyle \sum_{\Delta R<1.0} E_{\mathrm{T}}^{\mathrm{jet}} > x$ & 
Sums the energy of jet TOBs with $E_{\mathrm{T}}\geq 15$~\gev\ and centre within a cone of radius $\Delta R=1.0$ around each jet~\cite{Achenbach:2008zzb,Lampl:2008};
jets considered in the energy sum are distinct from the jet defining the cone centre. The threshold ($x$) for the energy sum is a parameter of the algorithm. \\
 
\noalign{\smallskip}\hline\noalign{\smallskip}
 
Late muon & Finds the highest-\pt{} muon in the next BC and combines it with the input lists associated with the current BC. This results in a tighter latency budget for this algorithm.  \\
 
\noalign{\smallskip}\hline\noalign{\smallskip}
\end{tabular*}
 
}
\end{center}
 
\label{tab:algo_def}
 
\end{table}

Examples of topological algorithms include angular requirements, invariant mass or transverse mass requirements, and global event requirements, with the flexibility to use central ($|\eta|<3.2$ for jets, $|\eta|<1.0$ for muons) and/or forward ($|\eta|>3.2$ for jets, $|\eta| > 1.0$ for muons) L1 trigger objects. Calorimeter and muon information can also be combined, for example to search for lepton-flavour-violating $B_{(s)}^{0}\rightarrow e^{\pm}\mu^{\mp}$ decays~\cite{Guadagnoli:2017jcl}. Also, L1 trigger objects can be required to come from different BCs, a requirement that can be useful in searching for heavy long-lived highly ionising particles that may be detected during the BC following the collision.
 
Angular selection criteria include requirements on the polar and azimuthal distances ($\Delta\eta$ and $\Delta\phi$)  between two or more trigger objects of the same or different type. These distances can be tailored to the kinematic properties of specific signal processes to select the objects satisfying the criteria or to discard a pair of overlapping objects from different lists.
Algorithms summing the energy of all the objects inside a cone of a certain radius (referred to as simple cone algorithms) are also available and used to seed HLT triggers for large-radius jets. They have proven to be helpful in analyses exploiting jet substructure.
 
Topological mass and event criteria include invariant and transverse masses ($m_{\mathrm{inv}}$ and $m_{\mathrm{T}}$), frequently used in many physics analyses, as well as event hardness ($H_{\mathrm{T}}$), defined as the scalar \pt{} sum of all jet TOBs in the event.
 
Window acceptance selections place geometrical requirements on the azimuthal and pseudorapidity positions of given trigger objects. They are most useful in the commissioning of new systems or testing of prototypes by providing specific triggers limited only to their acceptance area, with only a marginal impact on the total ATLAS L1 budget rate. This has been used, for example, to study the hardware Fast Tracker~\cite{Shochet:1552953} system partly installed during Run 2 and to commission the demonstrator set-up installed to verify the full functionality of the new Liquid Argon trigger readout system for the Phase-I upgrade~\cite{CERN-LHCC-2011-012}.
 
\FloatBarrier
\section{Simulation and validation}
\label{sec:validation}
The logic of the topological algorithms and their implementation in the hardware is validated
at various levels, before and during data-taking. Before data-taking operations,
the firmware is simulated in VHDL and
basic checks of the internal logic consistency are performed. Well-defined input data are processed
through the hardware via a playback mechanism and the decisions are examined. All L1Topo algorithms
are also simulated and coded within the L1 and HLT software framework as well as stand-alone.
This allows the validation and optimisation of the algorithms in terms of expected trigger rates
and signal acceptances.
 
The L1Topo system and its integration with the other ATLAS systems is tested and validated in situ
outside data-taking periods. In these tests, `hot towers' are generated in the trigger system by
overwriting the normal digitisation parameters of selected channels in the L1 calorimeter trigger
electronics with an alternative calibration that converts pedestal fluctuations, which would normally be removed
by a noise cut, into a high-\pt{} signal. These are artificially formed in
predefined regions of the subdetectors and issue a trigger that is used
as input to the L1Topo hardware. By knowing the exact position of the regions where the trigger originated, L1Topo
algorithms requiring a given angular position can be cross-checked. The time of
the arrival of the decisions is also checked to ensure that they are all well aligned with
the triggered events.
 
Once the firmware is validated and the L1Topo system well aligned in time, it is deployed
for data-taking. During data-taking, the system is constantly monitored. All
L1Topo algorithms are simulated in the HLT software running online in the HLT computing farm.
A fraction of the L1 accepted events are processed in real time through this simulation at
the HLT. This allows the comparison of the L1Topo hardware decisions and the simulated
decisions both during and after data-taking. Both the statistical and event-by-event differences
are displayed in various histograms for online monitoring~\cite{Aad:2020wji}. To achieve a precise
comparison, the algorithms are simulated using integers instead of floating-point precision for the TOBs properties,
as used in the L1Topo hardware. However, the exact implementation of the hardware logic in
software is difficult to achieve and differences between simulation and hardware decisions
at rates of a few per mille for less than $10\%$ of the L1Topo algorithms were observed for the following reasons.
Most topological algorithms use lists of TOBs ordered by decreasing \pt.
In cases where the TOBs have identical \pt, there is ambiguity in the calculation of the kinematical
variables that use them, such as invariant masses or differences in object position, potentially resulting in different
trigger decisions. These ambiguities were fully simulated only for the case of muon TOBs.
In addition, the hardware decisions take into account
overflow conditions, which are not simulated. The rate of overflows during the $pp$ data-taking was
at most a few Hz.
 
\FloatBarrier
\section{Trigger performance}
\label{sec:performance}
By offering a large variety of kinematic selections, the topological triggers can introduce requirements at L1 that are inspired by offline physics analyses. This approach results in significantly improved background event rejection and improved acceptance of physics signal events for many analyses, allowing the strict constraints of the L1 rate to be met, despite the increase in the instantaneous luminosity.
This section highlights the performance of the topological triggers for a few chosen physics signatures.
 
\subsection{$B$-physics}
 
\begin{figure}
\centering
\subfloat{
\includegraphics[height=0.30\textwidth]{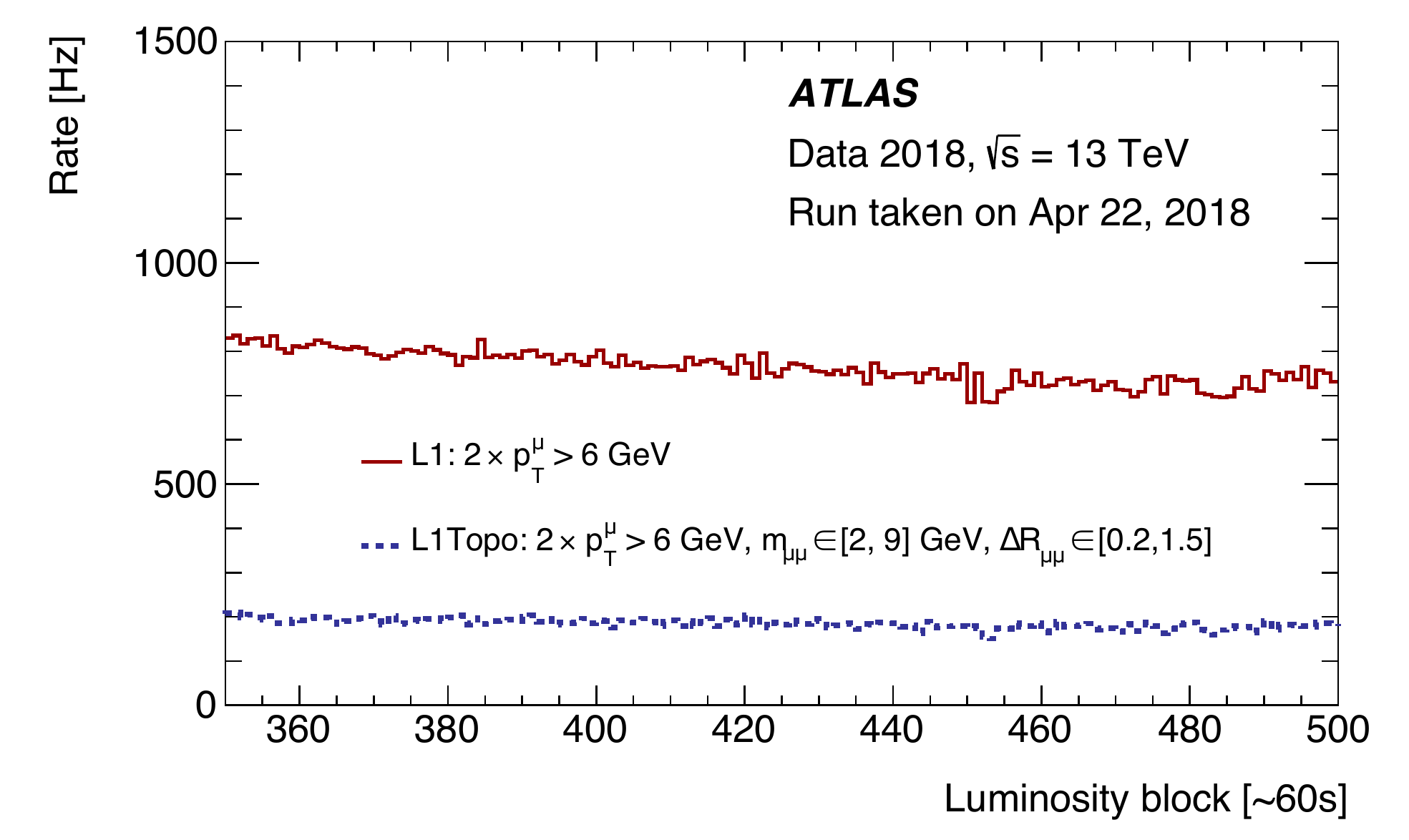}
\label{fig:Jpsimumu-rate}
}
\subfloat{
\includegraphics[height=0.30\textwidth]{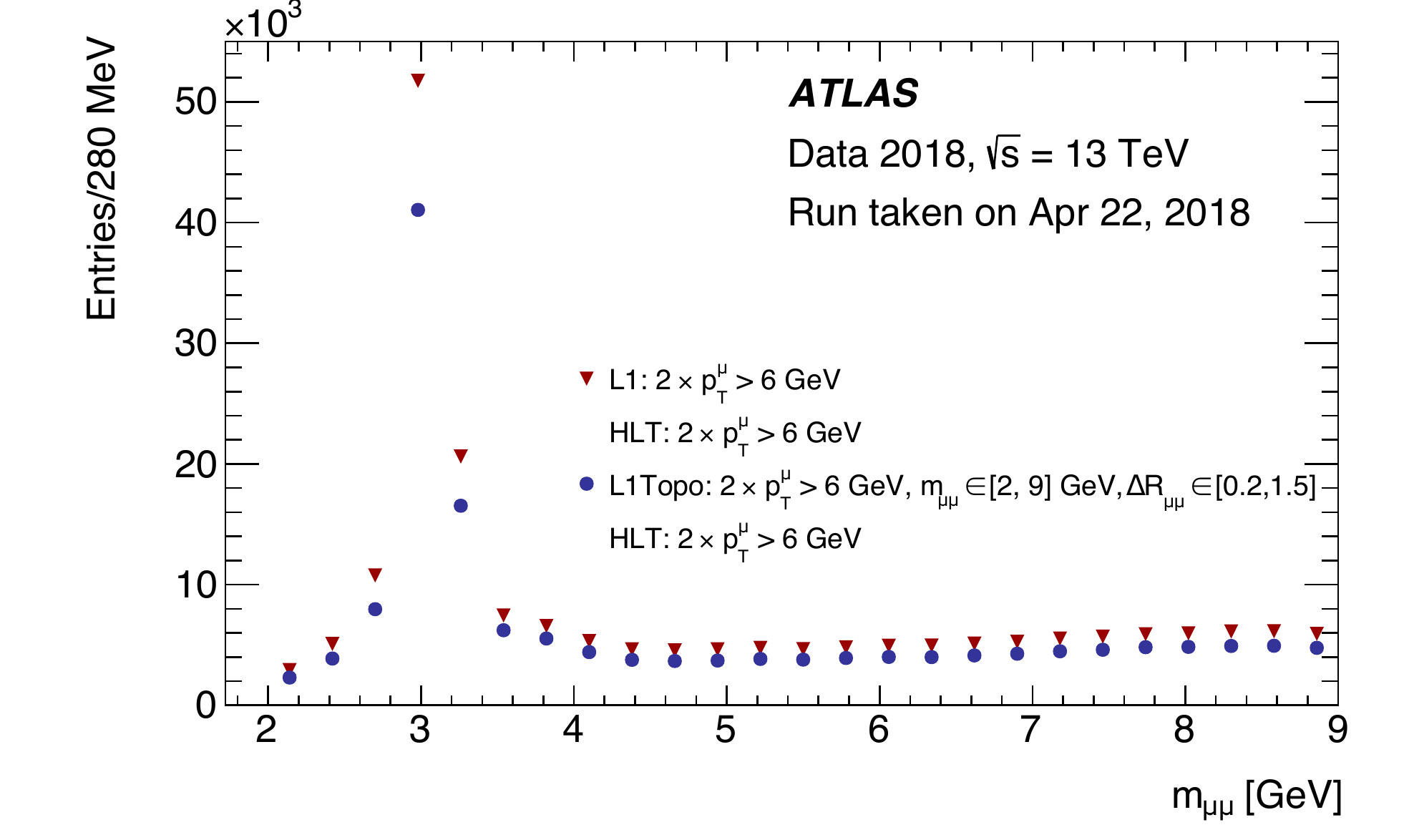}
\label{fig:Jpsimumu-eff}
}
\caption{Comparison of L1 dimuon triggers for muons with $\pt > 6$~\GeV\ with and without topological requirements: \protect\subref{fig:Jpsimumu-rate} L1 accept rate versus luminosity block number (one luminosity block is the time interval of data recording over which the experimental conditions are assumed to be constant, usually one minute) for an LHC $pp$ fill with peak luminosity of $2.6 \times 10^{33}$~cm$^{-2}$s$^{-1}$; \protect\subref{fig:Jpsimumu-eff}~invariant mass spectrum of the offline dimuon pair in events selected by the L1 and HLT dimuon triggers.}
\end{figure}

The first example is taken from the ATLAS $B$-physics programme, which strongly benefits from the use of L1Topo triggers. Many of these analyses are based on the identification of $B$ and $J/\psi$ mesons via their decay products, including  very low \pt{} electron and muon pairs. For example, for the $J/\psi\to\mu\mu$ selection~\cite{Aad:2020jfw}, the non-topological trigger requires two muons with \pt{} above 6~\GeV.  As shown in Figure~\ref{fig:Jpsimumu-rate}, this trigger yields a L1 accept rate of up to 0.8~kHz for a luminosity of ${\sim}1.9\times 10^{33}$~cm$^{-2}$s$^{-1}$. However,  by introducing topological requirements at trigger level that are closer to the actual selections in the offline analysis, such as requiring the angular separation between the two muon TOBs to be in the range $0.2 \leq \Delta R(\mu_1,\mu_2) \leq 1.5$ and the invariant mass of the dimuon pair ($m(\mu_1,\mu_2)$) to be in the range 2--9~\GeV, the trigger rate is reduced by a factor of four, as shown in Figure~\ref{fig:Jpsimumu-rate}.  This comes with a loss in signal efficiency of about 20\%, due to signal acceptance for the $\Delta R(\mu_1,\mu_2)$ selection, and, even more importantly, without introducing any bias or distortion in the mass distribution, as shown in Figure~\ref{fig:Jpsimumu-eff}. Reaching a sustainable trigger rate while keeping the muon \pt{} threshold at the same value, 6~\GeV, as in the corresponding non-topological trigger of Run~1 (2009--2013)~\cite{Aad:2014sca} was crucial for these analyses. An increase in the \pt{} threshold values would have significantly cut into the signal acceptance.
A similar strategy is followed for triggering $J/\psi \rightarrow ee$ events, used for example to study the performance of
low-$E_{\text{T}}$ electrons. Prescaled triggers requiring at least one EM TOB with $\et > 3$~\GeV\ and an EM TOB pair with an invariant mass
between 1 and 5~\GeV\ are used to reduce the rate and efficiently select these events~\cite{Aad:2019wsl}.
 
\begin{figure}
\centering
\subfloat{
\includegraphics[height=0.335\textwidth]{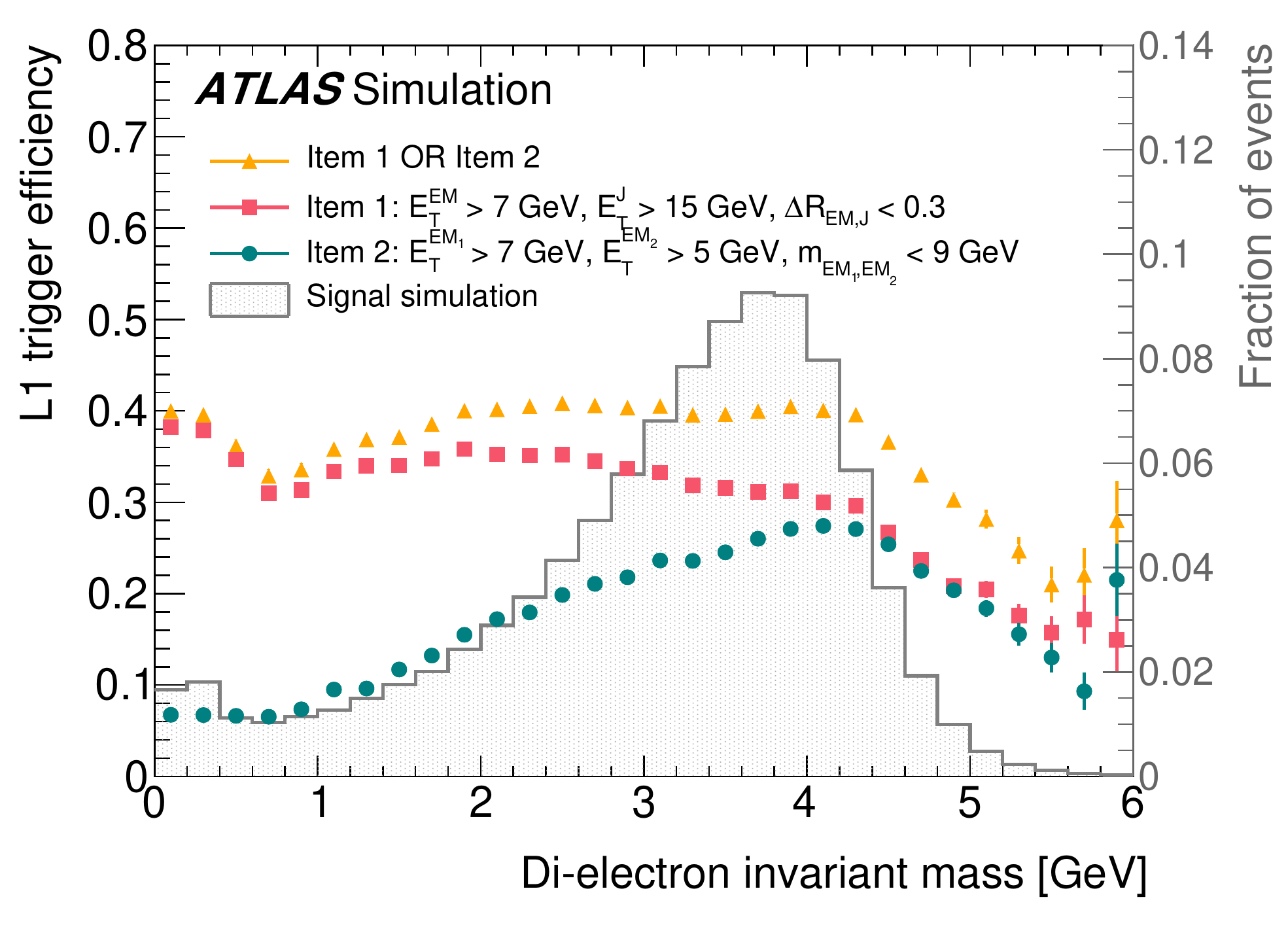}
\label{fig:kee-eff}
}
\hspace{0.2cm}
\subfloat{
\includegraphics[height=0.335\textwidth]{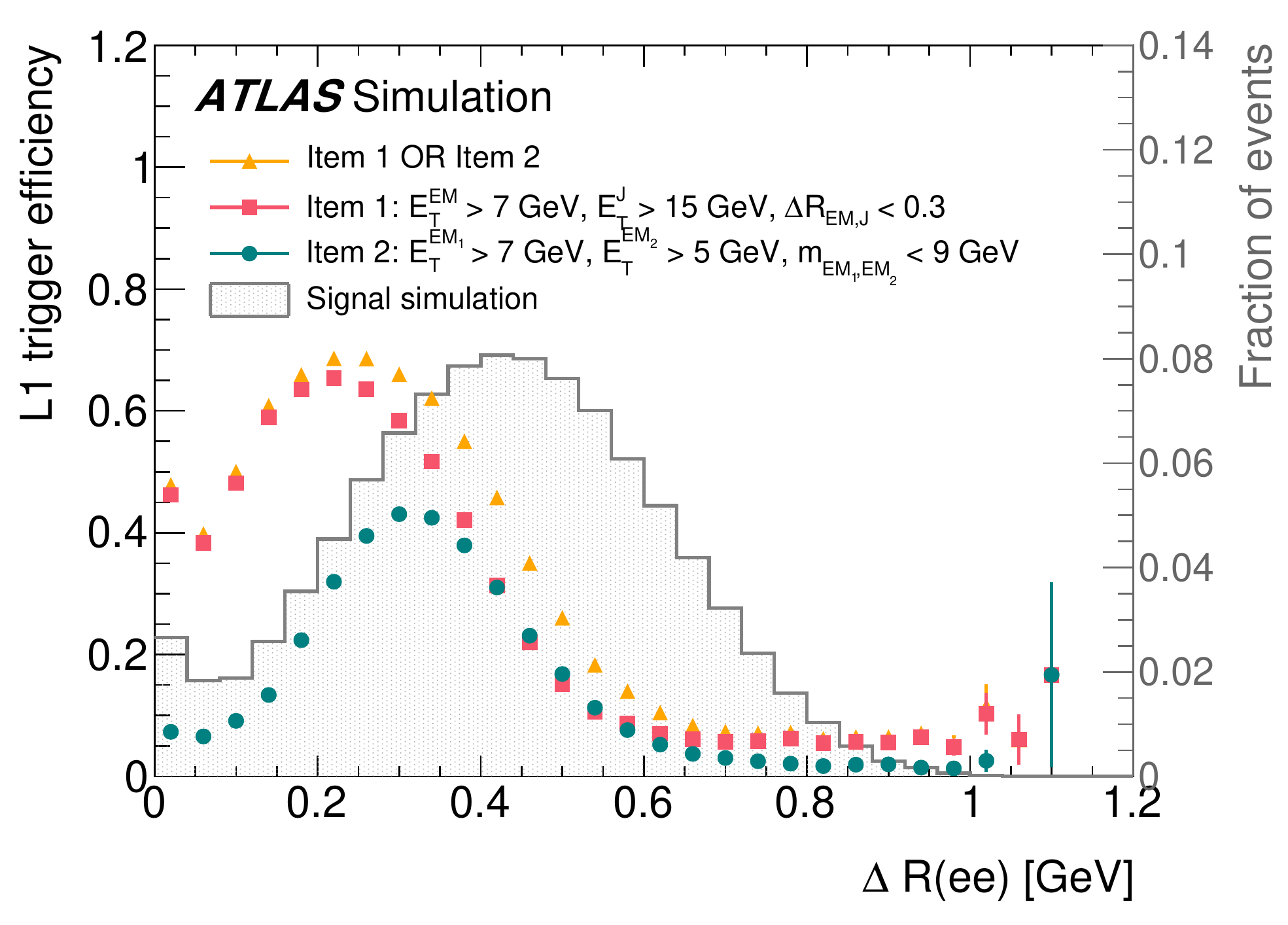}
\label{fig:kee-eff_2}
}
\caption{ Efficiency of L1Topo triggers developed for $B$-meson decays with di-electron final states. The efficiency is determined using non-resonant
$B^{0}_{d} \rightarrow  K^{*0} e^{+} e^{-}$ decays, and it is shown \protect\subref{fig:kee-eff}~as a function of the offline invariant mass of the signal electron pairs and  \protect\subref{fig:kee-eff_2}~as a function of their angular separation $\Delta R$. The L1 efficiency (coloured filled markers) is displayed separately for the two considered
L1 items as well as their combination. The expected signal distribution is superimposed (filled grey histogram).}
\end{figure}

A second interesting L1Topo use case from another $B$-physics measurement is the test of lepton-flavour universality by means of the $R_{K^{*0}}$
asymmetry analysis.
The experimental signature~\cite{Aaij:2017vbb} is characterised by the presence of a neutral kaon ($K^{*0}$) and a pair of opposite-sign collimated electrons or muons produced via radiative electroweak interactions. The main challenge in triggering on this signature at the LHC is the expected \pt{} of the leptons, which can be as low as a few \GeV, approximately consistent with the $B^{0}$--$K^{*0}$ mass difference plus the $B^{0}$ \pt.
This signature is especially challenging in the electron channel, where the background rate at L1 is much higher than in the muon channel.
The developed trigger strategy in the electron channel relies on the L1Topo system combining two algorithms. The first
algorithm triggers on electron pairs (EM$_{1}$, EM$_{2}$) having \pt{} thresholds of 7 and 5~\GeV\, respectively,
and invariant mass below 9~\GeV.
A second algorithm targets a topology where the electrons partially or fully overlap with each other;
for such cases the energy of the electrons is not fully contained in the narrow $\Delta\eta\times\Delta\phi=0.2\times 0.2$
geometrical L1 electron size. In order to fully measure the momenta of the electrons, thus minimising the impact of energy deposit loss, a wider jet (J) with $\Delta\eta\times\Delta\phi=0.8\times 0.8$ and $\pt > 15$~\GeV\ is required to overlap, based on $\Delta R$, with an electron (EM) with $\pt \geq 15$~\GeV.
These two algorithms are combined in a logical OR. As shown in Figure~\ref{fig:kee-eff},
the combination of the two L1Topo trigger items yields a L1 efficiency of about $40\%$ with limited variation across the bulk of the expected signal's invariant mass distribution.
The maximum L1 efficiency as a function of $\Delta R$ is about $70\%$, as seen in Figure~\ref{fig:kee-eff_2};
the loss of efficiency for $\Delta R>0.4$ is driven by the L1 \pt{} thresholds.
The combination of these two L1Topo algorithms helps to reject background events.
In order to limit their rate at high luminosity, single-muon and dimuon regions-of-interest were required in the same event, with the muons expected to originate from a second $b$-hadron in the event.

\subsection{Higgs physics}
 
\begin{figure}[htb]
\centering
\subfloat{
\includegraphics[height=0.39\textwidth]{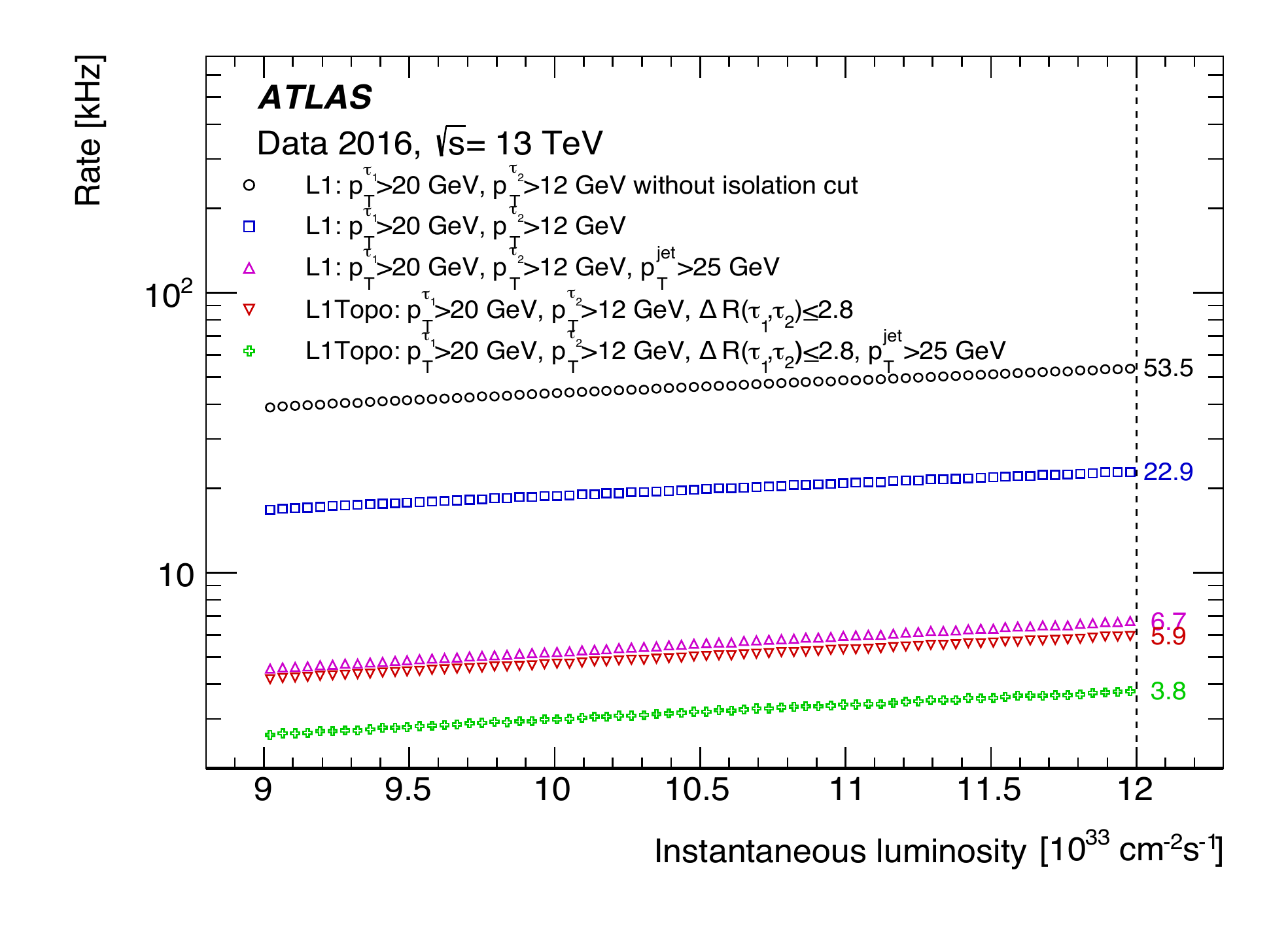}
\label{fig:Htautau-rate}
}
\subfloat{
\includegraphics[height=0.39\textwidth]{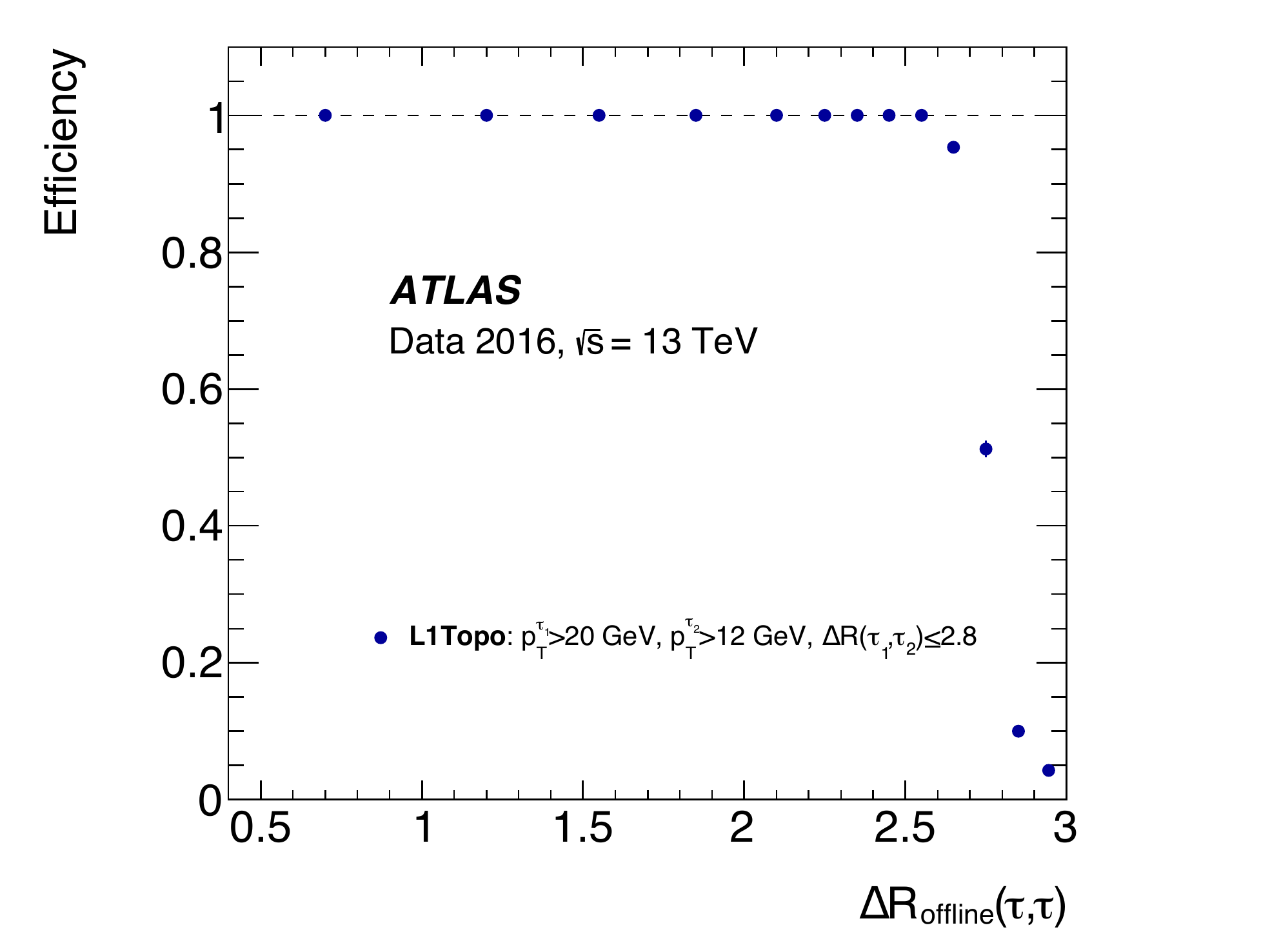}
\label{fig:Htautau-eff}
}
\caption{  \protect\subref{fig:Htautau-rate} Comparison of the L1 rate as a function of the instantaneous luminosity for di-\tauhad{} triggers
with (open red triangles) and without (open blue boxes) topological requirements;
the same figure also shows the L1 rate as a function of the instantaneous luminosity for di-\tauhad{} triggers with (open green crosses) or without (open pink triangles) the requirement of an additional jet; the L1 rates corresponding to an instantaneous luminosity of $12\times10^{33}$~cm$^{-2}$s$^{-1}$ are also displayed.
\protect\subref{fig:Htautau-eff}
Efficiency of the topological di-\tauhad{} trigger for events passing the non-topological trigger as a function of the $\Delta R(\tau_1,\tau_2)$ between the two $\tau$-leptons.
Algorithms requiring an additional jet also apply the selection $\Delta R(\tau, j) >0.1$ in order to prevent double counting of tau and jet TOBs.}
\end{figure}

Another application of L1Topo triggers is the measurement of Higgs boson decays into two hadronically decaying $\tau$-leptons, $H\to\tauhad\tauhad$ \cite{ATLAS-CONF-2017-061}. This analysis relies on a di-\tauhad{} trigger requiring two isolated tau TOBs passing relatively low \pt thresholds of 20~\GeV\ and 12~\GeV\ at L1. Keeping the \pt thresholds as low as possible is very important for sensitive measurements of the Higgs boson's coupling to fermions.
This can be achieved either by requiring one additional jet with $\pt >25$~\GeV\ in the event, which originated from vector-boson fusion produced Higgs bosons or quantum chromodynamics initial-state radiation for example, or by using a topological trigger that introduces an additional requirement on the angle between the two $\tau$ objects, $\Delta R(\tau_1,\tau_2) \le 2.8$, to help reject a large fraction of the multi-jet background events while retaining most of the signal events.
As observed in Figure~\ref{fig:Htautau-rate}, the additional jet required at L1 reduces the trigger rate by a factor of four without raising the tau TOB \pt{} thresholds. The reduction is similar to the rate reduction from the topological trigger requirement mentioned above.
A further reduction by a factor of 1.5 is achieved when adding a second topological requirement, namely the disambiguation algorithm that requires a jet candidate with $\pt >25$~\GeV\ and with $\Delta R(\tau, j) > 0.1$ with respect to each of the \tauhad{} candidates in the selected pair.
Figure~\ref{fig:Htautau-eff} shows the efficiency of one topological trigger relative to the non-topological di-\tauhad{} trigger. Thanks to the good angular resolution of the tau TOBs, the trigger efficiency turn-on is very sharp in $\Delta R$, making the topological trigger fully efficient for an offline selection of $\Delta R(\tau_1,\tau_2) \leq$ 2.6. Typically, a requirement of $\Delta R(\tau_1,\tau_2) \leq$~2.4 is applied in the offline analysis, independently of the use of the topological trigger. Therefore, this topological algorithm does not affect the signal acceptance in the analysed phase space. These L1Topo triggers were operated without a prescale factor in the Run~2 data-taking.
 
\begin{figure}[htb]
\centering
\subfloat{
\includegraphics[height=0.41\textwidth]{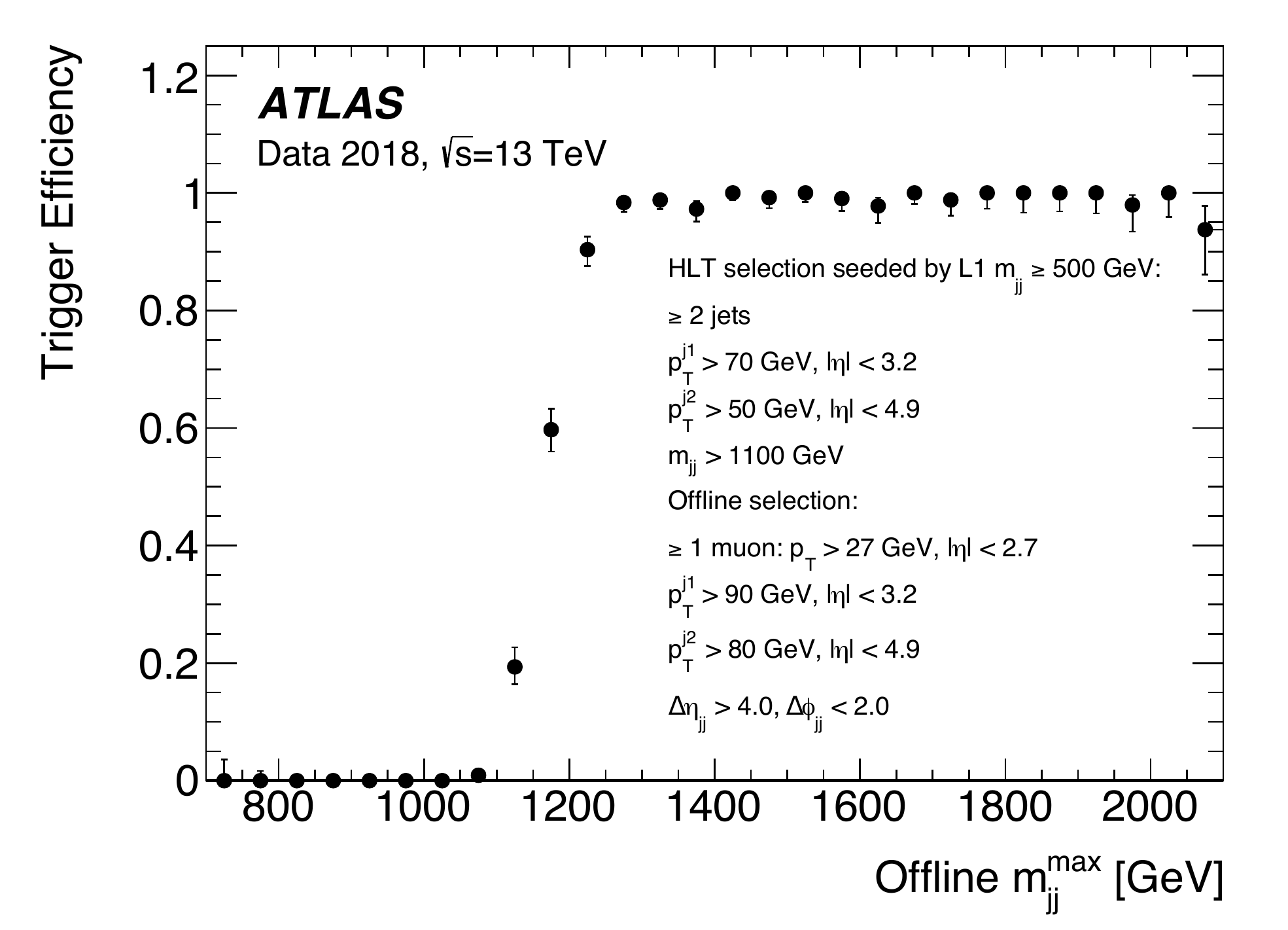}
\label{fig:mjj-nff}
}
\subfloat{
\includegraphics[height=0.41\textwidth]{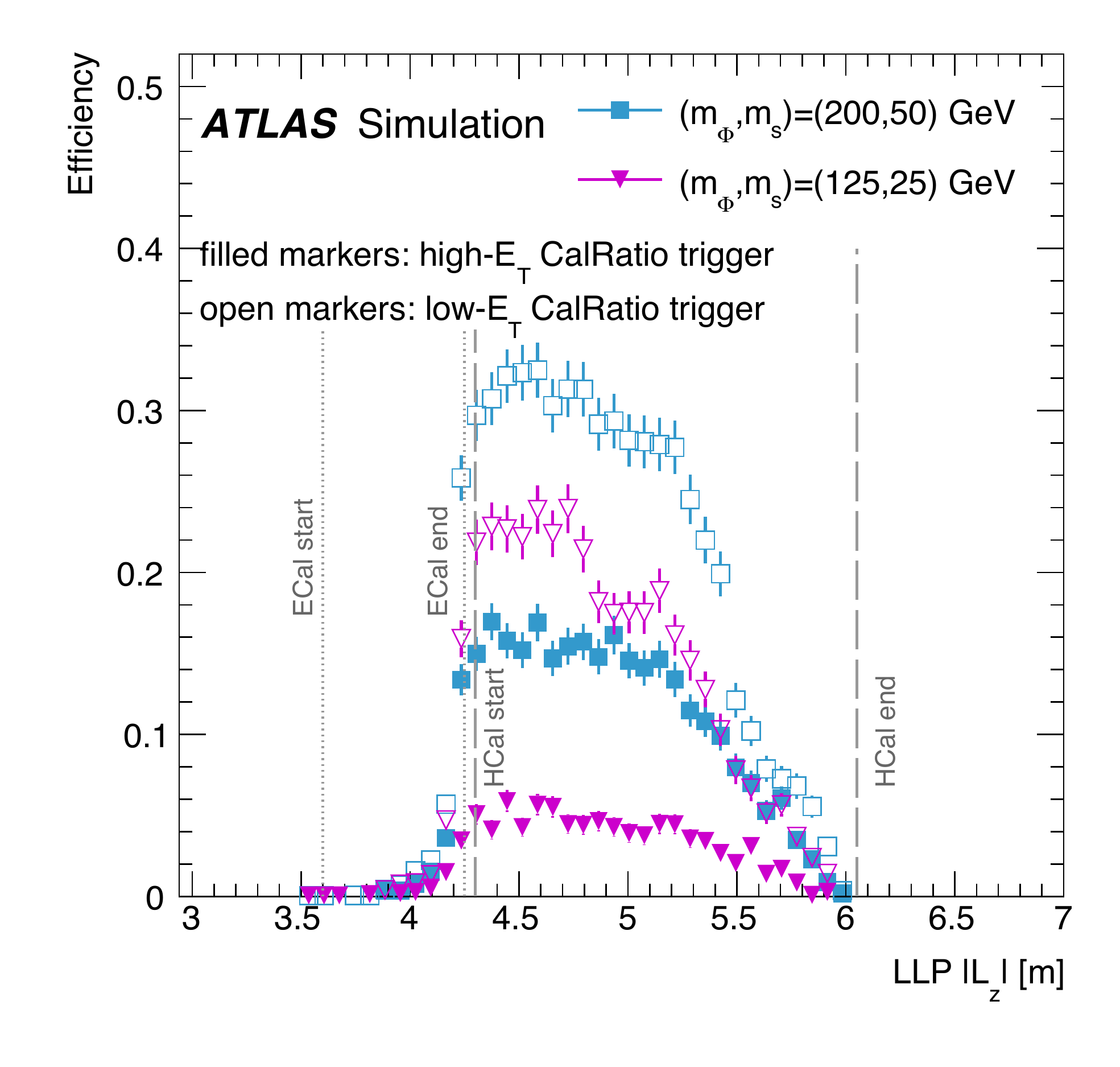}
\label{fig:llp}
}
\caption{\protect\subref{fig:mjj-nff} Efficiency of the HLT dijet trigger as a function of the maximum offline dijet mass.
The L1 topological algorithm used in this trigger requires the dijet invariant mass to be above 500~\GeV\ and at least one of the two jets to be in the central part of the detector ($|\eta|<$~3.2). The efficiency is measured using events that are selected using a single muon trigger with $\pt >$~27~\GeV. In addition, events are required to have at least two jets with $\et >80$ and 90~\GeV, respectively. Angular requirements between the jets are also applied offline.
\protect\subref{fig:llp} Efficiency of the high-\et{} non-topological CalRatio trigger (filled markers) and of the low-\et{} topological CalRatio trigger (open markers) as a function of the absolute value of the LLP longitudinal decay position, $L_z$. Different mass hypotheses for the signal models are represented in different colours. The grey dotted (dashed) lines show the start and end $z$-positions of the electromagnetic (hadronic) calorimeter.
}
\end{figure}

Topological triggers are also useful for other Higgs analyses.
For example, event hardness and invariant mass selections can be used to trigger events in which the Higgs boson is produced via vector-boson fusion. By trying to identify the two jets from the scattering process 
the signal acceptance of Higgs boson decays to invisible particles can be enhanced to about $30\%$ with respect to using the non-topological missing transverse energy requirement only.
The trigger is designed to select a first jet in the central part of the detector ($|\eta|<3.2$) and a second jet in the full detector acceptance range ($|\eta|<4.9$) forming an invariant mass ($m_{jj}$) greater than 500~\GeV.
The pseudorapidity restriction for one of the two jets allows the rate to be reduced by a factor of five compared to using an invariant mass requirement considering all jet combinations, at the price of a 10\% signal loss.
Additional requirements on the pseudorapidity difference and azimuthal separation between the two jets, $\Delta\eta_{jj}>4.0$ and $\Delta\phi_{jj}<2.0$, are added to the HLT to further reject the multi-jet background. Figure~\ref{fig:mjj-nff} shows the trigger efficiency of the HLT trigger item seeded by the topological algorithm as a function of the maximum offline dijet mass, $m_{jj}^{\mathrm{max}}$;
the HLT trigger requires $m_{jj}>1100$~\gev, a first jet with \pt$>70$~\gev\ and $|\eta|<3.2$, and a second jet with \pt$>50$~\gev\ and $|\eta|<4.9$.
The efficiency is measured using events with at least one offline-reconstructed muon with $\pT > 27$~\GeV.
In addition, the events are required to have at least two anti-$k_t$ jets~\cite{Fastjet,Cacciari:2008gp}
with a radius parameter $R=0.4$, where one jet has $\pt >90$~\GeV\ and $|\eta| < 3.2$, while another jet has $\pt >80$~\GeV. The $\Delta\eta_{jj}$ and $\Delta\phi_{jj}$ angular requirements are also applied offline,
as in the trigger requirement.
The efficiency of the HLT trigger seeded by the L1Topo algorithm to select events passing the offline analysis cuts is close to $100\%$ for events
with $m_{jj}^{\mathrm{max}}>1300$~\gev.
 
\subsection{Long-lived particles}
 
The new capabilities of the L1Topo system also allow the exploration of new ideas for unconventional signatures, such as the detection of long-lived particles (LLP).
One example where L1Topo triggers are successfully used is an analysis that probes Hidden Sector models predicting decays of the Higgs boson or a new heavy neutral scalar particle $\Phi$ into neutral LLPs \cite{Aaboud:2019opc, TRIG-2012-02}.
Decays of neutral LLPs in the outer layers of the electromagnetic calorimeter or in the hadronic calorimeter results in topologies characterised by displaced collimated jets with little energy deposited in the electromagnetic calorimeter and, therefore, a large hadronic to electromagnetic energy ratio, $E_{\mathrm{HAD}}/E_{\mathrm{EM}}$. To target these events, a L1 trigger, called high-\et{} CalRatio, was designed to select narrow $\Delta\eta\times\Delta\phi=0.2\times 0.2$ jets with transverse energy $\et >60$~\GeV\ in a combined region of the electromagnetic and hadronic calorimeters. The \et{} threshold is driven by the high rates due to the pile-up conditions in Run~2; however, it strongly limits the efficiency of detecting signal events for models with $m_\Phi$ below 200~\GeV.
This threshold can be lowered to 30~\GeV\ by configuring a dedicated trigger, called low-\et{} CalRatio, that exploits the specific topology of these events.
Considering that these jets are expected to be very collimated (with most of the energy contained in a cone of $\Delta R =0.1$ around the jet axis), it is better to identify their calorimeter energy deposits using tau TOBs instead of jet TOBs, given their smaller area~\cite{Achenbach:2008zzb}.
The low-\et{} CalRatio trigger requires the leading-\et{} tau TOB to be above 30~\GeV\ and not to overlap with any EM TOBs with $\et >3$~\GeV. The same requirements are also applied to the second-leading tau TOB if present. Given that the tau TOBs are reconstructed using both the electromagnetic and hadronic calorimeters while the EM TOBs are reconstructed only with the electromagnetic calorimeter, the above selection is equivalent to requiring an object with high $E_{\mathrm{HAD}}/E_{\mathrm{EM}}$ ratio and relies on geometrically overlapping EM and tau TOBs in the $\eta$--$\phi$ plane.  This rejects a large fraction of the background events.
Figure \ref{fig:llp} shows the efficiency of the high-\et{} and low-\et{} CalRatio triggers as a function of the LLP longitudinal decay position. The topological algorithm recovers up to about $20\%$ of trigger efficiency for low-$m_\Phi$ hypotheses.
 
The capability of L1Topo to trigger on muon TOBs detected in the BC after the BC of interest is also exploited. These triggers combine the requirement of one TOB present in a given BC with the presence of a muon TOB in the following BC. They are particularly useful in searches for long-lived particles.
One such example consists of requiring a jet candidate with a \pt{} above 50~\GeV\ in the current BC together with a muon candidate with a \pt{} above 10~\GeV\ in the following BC. The much reduced muon \pt{} threshold together with the also relatively low jet \pt{} threshold allows the efficiency to be improved by a factor of 2--3 in searches for long-lived highly ionising particles with masses above 500~\GeV, with electric charge greater than one elementary charge, and with no strong interaction.
 
\subsection{Large-$R$ jets}

\begin{figure}
\centering
\subfloat{
\includegraphics[height=0.33\textwidth]{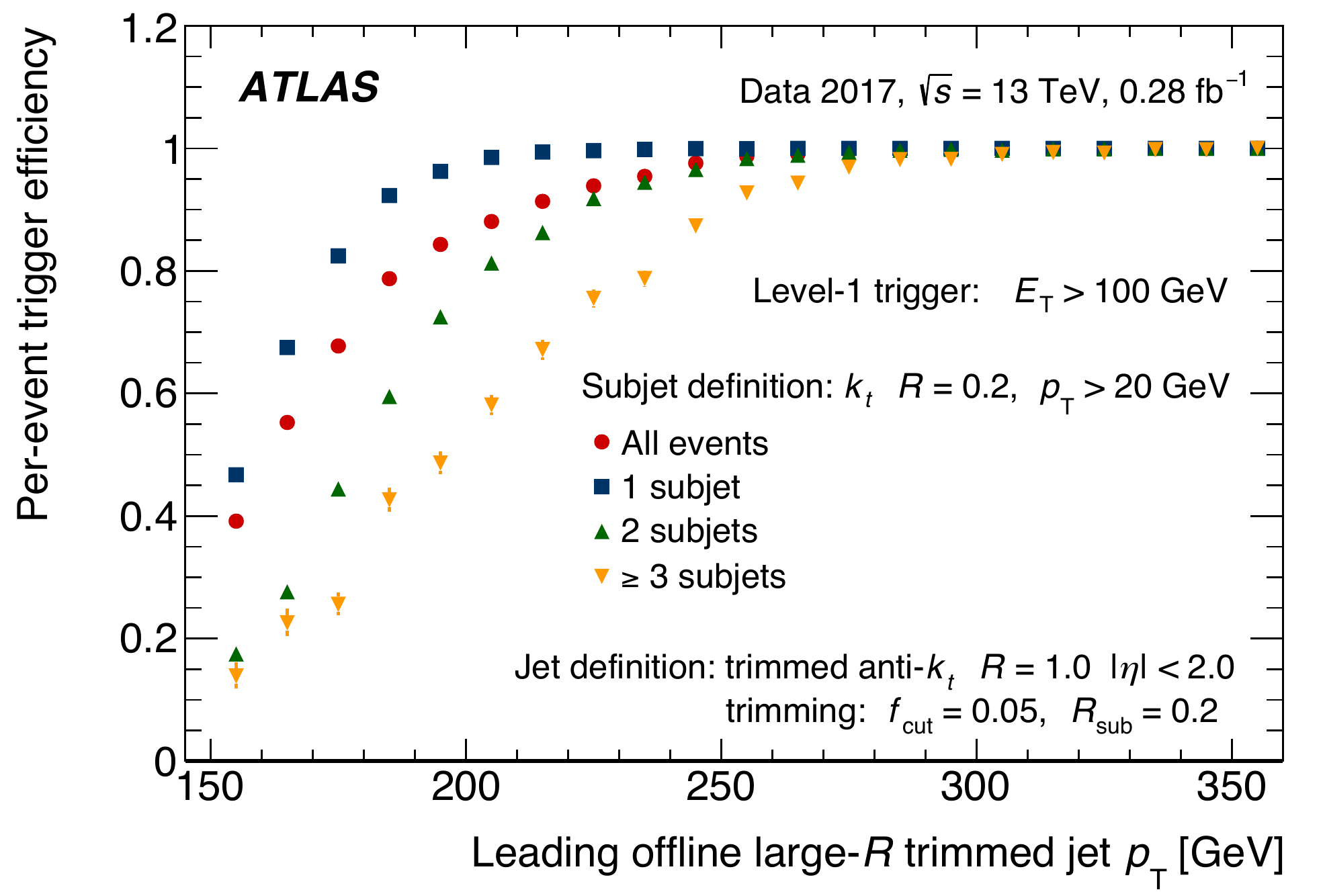}
\label{fig:beforeSC}
}
\subfloat{
\includegraphics[height=0.33\textwidth]{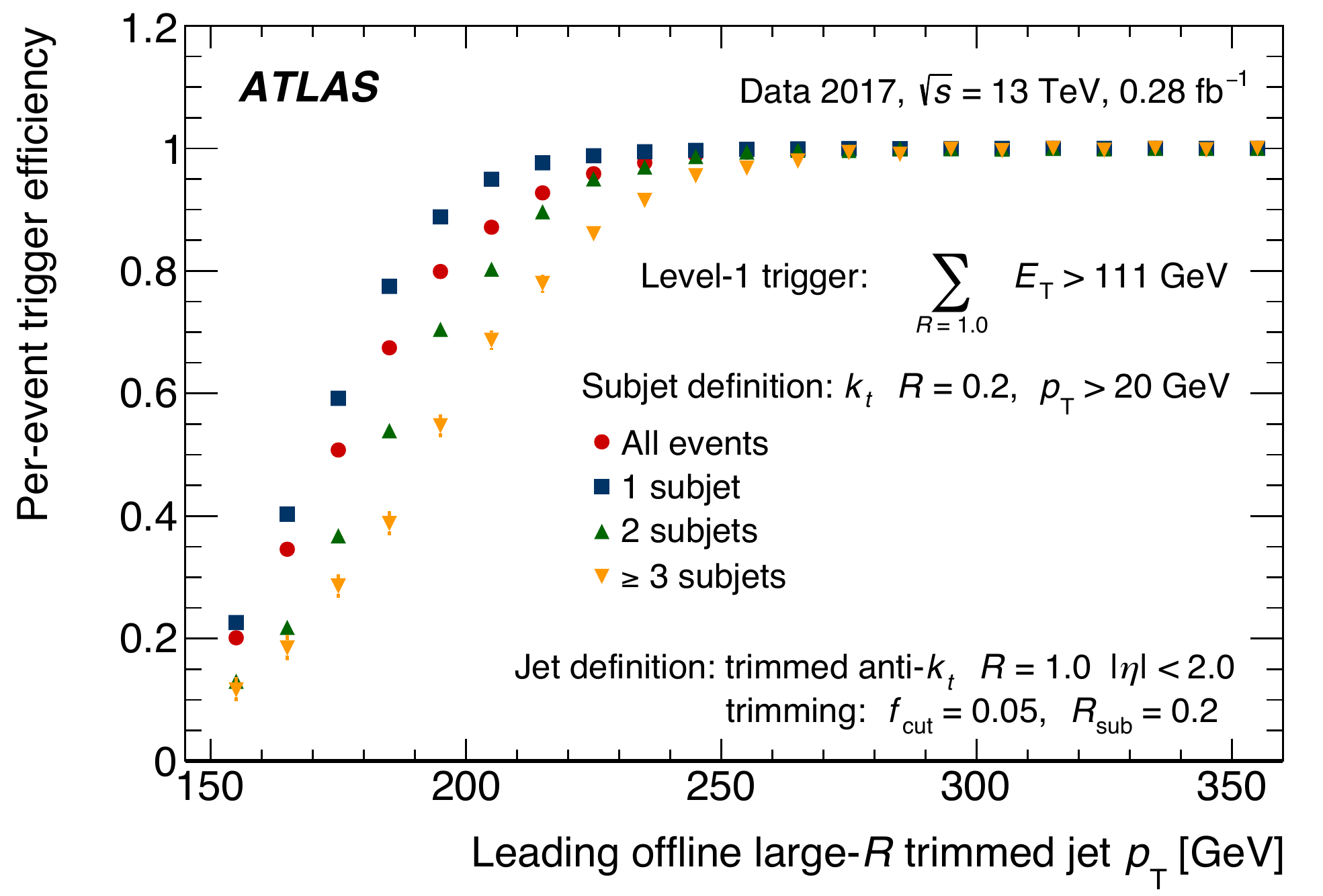}
\label{fig:afterSC}
}
\caption{Trigger efficiency curves as a function of the leading offline trimmed jet \pt \cite{Krohn_2010} for events with jets of one, two or more than two subjets selected using \protect\subref{fig:beforeSC} the standard jet sliding-window algorithm and \protect\subref{fig:afterSC} the L1Topo simple cone algorithm.  }
\end{figure}

Non-topological L1 jet triggers use a fixed $\Delta\eta\times\Delta\phi$ sliding-window method~\cite{Achenbach:2008zzb,Lampl:2008} with size $0.8\times 0.8$. These, however, fail to capture all the energy of jets with radius $R$ larger than 0.4.
The performance of this method deteriorates as the number of subjets within a jet grows, as it becomes more likely that a significant fraction of the jet energy falls outside of the selected $0.8 \times 0.8$ window.
Figure~\ref{fig:beforeSC} shows how an increase in the number of subjets reduces the trigger efficiency.
The use of L1Topo provides more efficient triggering for these large-$R$ jets.
The L1Topo simple cone algorithm is designed to sum the \et{} of all $0.8\times 0.8$ jet TOBs with $\et >15$~\GeV\ and a centre within a cone of $\Delta R =1$. The obtained \et{} sum, representing the energy of the large-$R$ jet, is required to be larger than 111~\GeV; this value is chosen to result in a trigger rate equals to that of the non-topological 100~\GeV\ jet trigger. Figure~\ref{fig:afterSC} shows the L1Topo simple cone algorithm trigger efficiency for events with jets of various numbers of subjets. The comparison of Figures~\ref{fig:beforeSC} and~\ref{fig:afterSC} shows that the L1Topo simple cone is able to
reduce by up to 30~\GeV\ the \pt\ value at which the trigger becomes fully efficient for events with jets of multiple subjets. It can also be observed that the L1Topo trigger efficiency for events with jets of one single subjet is worse than that corresponding to the single jet trigger. This is due to the increase in the jet \et\ requirement from 100~\GeV\ to 111~\GeV.
However, this small increase in threshold does not affect the offline selection efficiency, as the HLT triggers of single jets require a much higher \et, above 400~\GeV, thus removing the events under the L1Topo trigger efficiency turn-on.
In contrast, the HLT triggers aiming at final states involving the production of multiple hadronically decaying massive particles (dibosons, $t\bar{t}$, etc.) can use mass cuts to strongly suppress multi-jet events, the dominant source of a high trigger rate, and the minimum jet \et\ requirement can thus be relaxed to 330~\GeV. This threshold is well above the point at which the L1Topo cone algorithm is fully efficient for jets with subjet multiplicities of three or more, while being at the edge of the efficiency plateau for the traditional $0.8\times 0.8$ L1 jet algorithm. In 2018, the L1Topo simple cone algorithm became the default L1 large-$R$ jet algorithm used to seed all HLT triggers targeting boosted hadronically decaying massive particles.

\FloatBarrier
\section{Conclusions}
\label{sec:conclusion}
The L1 topological processor provides the ATLAS L1 trigger system with the capability of applying kinematic selections among muon and calorimeter-based trigger objects to substantially reduce the L1-accept trigger rate, while maintaining high signal acceptance, at instantaneous luminosities up to 2.1$\times$10$^{34}$~cm$^{-2}$s$^{-1}$. This new capability became available for the ATLAS physics programme during LHC Run~2.
The topological triggers are particularly important for physics analyses making use of low-\pt{} objects,
such as in Higgs boson and $B$-physics measurements.
The L1Topo system also makes it possible to trigger on objects from consecutive BCs.
The system was successfully installed and commissioned in 2016 and operated during $pp$ physics data-taking in 2017 and 2018. Simulation, validation and monitoring tools have been crucial for the commissioning and monitoring of the system.
 
\section*{Acknowledgements}  

We thank CERN for the very successful operation of the LHC, as well as the
support staff from our institutions without whom ATLAS could not be
operated efficiently.
 
We acknowledge the support of ANPCyT, Argentina; YerPhI, Armenia; ARC, Australia; BMWFW and FWF, Austria; ANAS, Azerbaijan; SSTC, Belarus; CNPq and FAPESP, Brazil; NSERC, NRC and CFI, Canada; CERN; ANID, Chile; CAS, MOST and NSFC, China; COLCIENCIAS, Colombia; MSMT CR, MPO CR and VSC CR, Czech Republic; DNRF and DNSRC, Denmark; IN2P3-CNRS and CEA-DRF/IRFU, France; SRNSFG, Georgia; BMBF, HGF and MPG, Germany; GSRT, Greece; RGC and Hong Kong SAR, China; ISF and Benoziyo Center, Israel; INFN, Italy; MEXT and JSPS, Japan; CNRST, Morocco; NWO, Netherlands; RCN, Norway; MNiSW and NCN, Poland; FCT, Portugal; MNE/IFA, Romania; JINR; MES of Russia and NRC KI, Russian Federation; MESTD, Serbia; MSSR, Slovakia; ARRS and MIZ\v{S}, Slovenia; DST/NRF, South Africa; MICINN, Spain; SRC and Wallenberg Foundation, Sweden; SERI, SNSF and Cantons of Bern and Geneva, Switzerland; MOST, Taiwan; TAEK, Turkey; STFC, United Kingdom; DOE and NSF, United States of America. In addition, individual groups and members have received support from BCKDF, CANARIE, Compute Canada, CRC and IVADO, Canada; Beijing Municipal Science \& Technology Commission, China; COST, ERC, ERDF, Horizon 2020 and Marie Sk{\l}odowska-Curie Actions, European Union; Investissements d'Avenir Labex, Investissements d'Avenir Idex and ANR, France; DFG and AvH Foundation, Germany; Herakleitos, Thales and Aristeia programmes co-financed by EU-ESF and the Greek NSRF, Greece; BSF-NSF and GIF, Israel; La Caixa Banking Foundation, CERCA Programme Generalitat de Catalunya and PROMETEO and GenT Programmes Generalitat Valenciana, Spain; G\"{o}ran Gustafssons Stiftelse, Sweden; The Royal Society and Leverhulme Trust, United Kingdom.
 
The crucial computing support from all WLCG partners is acknowledged gratefully, in particular from CERN, the ATLAS Tier-1 facilities at TRIUMF (Canada), NDGF (Denmark, Norway, Sweden), CC-IN2P3 (France), KIT/GridKA (Germany), INFN-CNAF (Italy), NL-T1 (Netherlands), PIC (Spain), ASGC (Taiwan), RAL (UK) and BNL (USA), the Tier-2 facilities worldwide and large non-WLCG resource providers. Major contributors of computing resources are listed in Ref.~\cite{ATL-SOFT-PUB-2020-001}.
 
 
\clearpage
\printbibliography

\clearpage
 
\begin{flushleft}
\hypersetup{urlcolor=black}
{\Large The ATLAS Collaboration}

\bigskip

\AtlasOrcid[0000-0002-6665-4934]{G.~Aad}$^\textrm{\scriptsize 101}$,    
\AtlasOrcid[0000-0002-5888-2734]{B.~Abbott}$^\textrm{\scriptsize 127}$,    
\AtlasOrcid[0000-0002-7248-3203]{D.C.~Abbott}$^\textrm{\scriptsize 102}$,    
\AtlasOrcid[0000-0002-2788-3822]{A.~Abed~Abud}$^\textrm{\scriptsize 36}$,    
\AtlasOrcid[0000-0002-1002-1652]{K.~Abeling}$^\textrm{\scriptsize 53}$,    
\AtlasOrcid[0000-0002-2987-4006]{D.K.~Abhayasinghe}$^\textrm{\scriptsize 93}$,    
\AtlasOrcid[0000-0002-8496-9294]{S.H.~Abidi}$^\textrm{\scriptsize 29}$,    
\AtlasOrcid[0000-0002-8279-9324]{O.S.~AbouZeid}$^\textrm{\scriptsize 40}$,    
\AtlasOrcid{N.L.~Abraham}$^\textrm{\scriptsize 155}$,    
\AtlasOrcid[0000-0001-5329-6640]{H.~Abramowicz}$^\textrm{\scriptsize 160}$
% ...
\bigskip
\\

% ...
$^{29}$Physics Department, Brookhaven National Laboratory, Upton NY; United States of America.\\
% ...
$^{36}$CERN, Geneva; Switzerland.\\
% ...
$^{40}$Niels Bohr Institute, University of Copenhagen, Copenhagen; Denmark.\\
% ...
$^{53}$II. Physikalisches Institut, Georg-August-Universit\"{a}t G\"ottingen, G\"ottingen; Germany.\\
% ...
$^{93}$Department of Physics, Royal Holloway University of London, Egham; United Kingdom.\\
% ...
$^{101}$CPPM, Aix-Marseille Universit\'e, CNRS/IN2P3, Marseille; France.\\
$^{102}$Department of Physics, University of Massachusetts, Amherst MA; United States of America.\\
% ...
$^{127}$Homer L. Dodge Department of Physics and Astronomy, University of Oklahoma, Norman OK; United States of America.\\
% ...
$^{155}$Department of Physics and Astronomy, University of Sussex, Brighton; United Kingdom.\\
% ...
$^{160}$Raymond and Beverly Sackler School of Physics and Astronomy, Tel Aviv University, Tel Aviv; Israel.\\
% ...

\end{flushleft}


\end{document}